\documentclass[Journal]{IEEEtran}
\IEEEoverridecommandlockouts
\usepackage{subfig} 
\usepackage{cite}
\usepackage{amsmath,amssymb,amsfonts}
\allowdisplaybreaks[4]
\usepackage{algorithmic,algorithm}
\usepackage{graphicx}
\usepackage{textcomp}
\usepackage{xcolor}
\usepackage{orcidlink}
\newtheorem{theorem}{Theorem}
\newtheorem{proposition}{Proposition}

\newtheorem{remark}{Remark}
\newtheorem{proof}{Proof}
\def\BibTeX{{\rm B\kern-.05em{\sc i\kern-.025em b}\kern-.08em
    T\kern-.1667em\lower.7ex\hbox{E}\kern-.125emX}}
\begin{document}

\title{Joint Subcarrier and Power Allocation for Localization-Oriented Multi-Carrier Underwater ISAC Systems\\
\thanks{This work was supported in part by the National Natural Science Foundation of China under Grant 42404001, 62271459, in part by Central University Basic Research Fund of China, Ocean University of China under Grant 202562022.}
}

\author{Wei Huang\orcidlink{0000-0002-8284-2310},~\IEEEmembership{Member,~IEEE,} Sitong Liu\orcidlink{0009-0006-9348-536X}, Christos Masouros\orcidlink{0000-0002-8259-6615},~\IEEEmembership{Fellow,~IEEE,} Hao Zhang\orcidlink{0000-0002-6454-0239},~\IEEEmembership{Senior Member,~IEEE,} T. Aaron Gulliver\orcidlink{0000-0001-9919-0323},~~\IEEEmembership{Senior Member,~IEEE,} Kaitao Meng*\orcidlink{0000-0001-7479-2280},~\IEEEmembership{Senior Member,~IEEE} 
	\thanks{Wei Huang, Sitong Liu and Hao Zhang are with the Faculty of Information Science and Engineering, Ocean University of China (email: hw@ouc.edu.cn; lst6811@stu.ouc.edu.cn; zhanghao@ouc.edu.cn).}
	\thanks{Christos Masouros is with the Department of Electrical and Electronic Engineering, University College London, London, UK (email: c.masouros@ucl.ac.uk).}
	\thanks{Kaitao Meng is with the Department of Electrical and Electronic Engineering, The University of Manchester, Manchester, UK. Corresponding author: Kaitao Meng (kaitao.meng@manchester.ac.uk).}
	\thanks{T. Aaron Gulliver is with the Department of Electrical and Computer Engineering, The University of Victoria, Victoria, Canada (email: agullive@uvic.ca).}
	\thanks{Manuscript received XX XX, 2026; revised XX XX, 2026.}}
\maketitle

\begin{abstract} 
Multi-carrier underwater integrated sensing and communication (ISAC) realizes concurrent acoustic communication and localization, yet the frequency selectivity of the acoustic channel creates uneven positioning capabilities across subcarriers. This paper proposes a localization-oriented subcarrier and power allocation scheme for multi-anchor underwater ISAC systems. We build a unified orthogonal subcarrier multiplexing framework and derive closed-form Cramér–Rao bounds (CRBs) under frequency-selective channels with colored noise. We derive a subcarrier-level localization information contribution from the time of arrival Fisher information, which enables localization-oriented resource allocation. We derive several theoretical properties regarding the distance and frequency dependence of the localization information metric, and establish a distance–frequency matching principle. These results provide a theoretical foundation for localization-oriented resource allocation and guide the design of a low-complexity greedy algorithm that improves positioning accuracy. Specifically, the algorithm iteratively allocates subcarriers to minimize the residual positioning-error objective, thereby improving positioning accuracy while satisfying communication quality-of-service (QoS) constraints. Simulations verify that our method yields much lower positioning error bound than uniform, random and high-frequency-first subcarrier allocation in symmetric or asymmetric topologies while guaranteeing communication QoS, achieving superior communication-localization tradeoffs.
\end{abstract}


\section{Introduction}
Accurate underwater localization and efficient communication are essential for underwater positioning, navigation, timing, and communication (PNTC) systems \cite{Chen2023PNT,Sun2023Loc,Draz2025LocCom}. Conventional underwater communication and localization are typically implemented separately, resulting in inefficient spectrum utilization. Integrated sensing and communication (ISAC) addresses this limitation by enabling communication and localization to share the same waveform, spectrum, and hardware resources \cite{Liu2020ISAC,Liu2023RaC,Shang2026Review}. By jointly supporting data transmission and positioning, ISAC improves spectral efficiency and resource utilization compared with conventional time-division approaches. Consequently, underwater ISAC has become a promising technology for next-generation underwater acoustic networks \cite{ICAL2022Review,Kwon2023ICAL}.

ISAC significantly improves spectrum efficiency by enabling communication and positioning to share the same acoustic resources \cite{Sturm2011Waveform,Karisan2011OFDM,Wang2023Beam}, but achieving high-accuracy localization in multi-carrier underwater systems remains challenging because underwater acoustic attenuation, ambient noise, and channel selectivity are strongly frequency-dependent, and different subcarriers provide heterogeneous ranging capabilities. 

The challenge becomes even more pronounced in multi-anchor underwater ISAC systems. Unlike single-link ranging, localization accuracy is jointly determined by the localization information contributed by all anchors as well as their geometric configuration \cite{Zhang2018Geo,Huang2024GDOP,Huang2025Optimal}. The localization information is a scalar metric derived from the Time-of-Arrival (TOA) Fisher information, which quantifies the positioning contribution of each subcarrier. Since the localization contribution of each anchor depends on both its assigned frequency resources and propagation conditions mentioned above, subcarrier allocation affects not only individual ranging accuracy but also the geometry of overall localization information. Therefore, frequency allocation in multi-anchor underwater ISAC is inherently a coupled resource allocation problem, where the optimal assignment must simultaneously account for frequency-dependent propagation characteristics and anchor cooperation.

Existing studies on underwater ISAC mainly focus on localization theory \cite{Islam2020Loc,Luo2021Loc,Huang2024Loc,Yan2025LocSurvey}, waveform design \cite{Sturm2011Waveform,Wang2023Beam,Lyu2023Wave}, and communication resource optimization \cite{Karisan2011OFDM,He2023Power,Dong2023Resource}. Most existing resource allocation schemes are communication-oriented and primarily optimize throughput, energy efficiency, or reliability. Such approaches are generally unsuitable for localization optimization because they neglect the heterogeneous localization information provided by different subcarriers under underwater acoustic propagation. Although localization-oriented resource allocation has attracted increasing attention in terrestrial radio-frequency (RF) cooperative localization systems \cite{Garcia2021LSPnet,Gao2024Locop,Singh2024Comop,Meng2025ISAC1,Meng2025ISAC2,Liu2026ComOp}, the unique frequency-dependent attenuation and propagation characteristics of underwater acoustic channels fundamentally alter the resource allocation mechanism. As a result, localization-oriented frequency allocation for multi-anchor underwater ISAC systems remains largely unexplored.

Unlike terrestrial wireless systems, underwater networks mainly rely on acoustic propagation due to severe electromagnetic attenuation in seawater. The absence of GNSS coverage makes accurate localization critical for
underwater missions. Moreover, the low propagation speed and frequency-dependent attenuation of acoustic channels lead to heterogeneous time-delay estimation performance among subcarriers. Therefore, effective localization-oriented frequency and power allocation in multi-carrier underwater ISAC systems remains challenging. In this work, we consider an AUV collecting data from multiple anchors while exploiting communication signals for localization. The objective is to improve positioning accuracy by optimizing orthogonal subcarrier allocation under communication constraints. The main
contributions are summarized as follows.

\begin{itemize}
\item We establish a unified multi-carrier underwater ISAC framework in which multiple anchor nodes deliver communication and localization services via orthogonal subcarrier allocation. Unlike existing communication-oriented resource management schemes, this framework targets high-precision positioning scenarios and explicitly characterizes the coupling among frequency allocation, power allocation, communication performance and localization accuracy, providing a theoretical foundation for localization-oriented ISAC design under communication constraints.

\item We derive closed-form CRBs for TOA estimation and target localization under frequency-selective underwater acoustic channels with colored ambient noise and interference. Based on these CRBs, a localization information metric is introduced to quantify the positioning contribution of individual subcarriers, enabling localization performance analysis directly in the frequency-allocation domain.

\item We formulate the subcarrier assignment problem as a communication-constrained A-optimal resource allocation problem that minimizes the positioning CRB while satisfying practical communication rate requirements. The resulting formulation explicitly exploits the heterogeneous localization capabilities of different subcarriers caused by frequency-dependent acoustic propagation and ambient noise characteristics.

\item We establish a distance-dependent frequency ordering theorem which indicates high frequency subcarriers should be assigned to close anchors and low frequency subcarriers to far anchors. This distance-frequency matching rule quantifies how propagation distance and carrier frequency jointly affect localization information, and provides direct theoretical guidance for localization-oriented subcarrier resource allocation.
\end{itemize}

The rest of this paper is organized as follows. Section \uppercase\expandafter{\romannumeral2} gives the system model. Section \uppercase\expandafter{\romannumeral3} provides the localization-oriented joint frequency and power allocation. Section \uppercase\expandafter{\romannumeral4} presents the greedy subcarrier allocation algorithm. Simulation results are given in Section \uppercase\expandafter{\romannumeral5}, and some concluding remarks are provided in Section \uppercase\expandafter{\romannumeral6}.

\section{System Model}

\begin{figure}[!htbp]
	\centering
	\includegraphics[width=\linewidth]{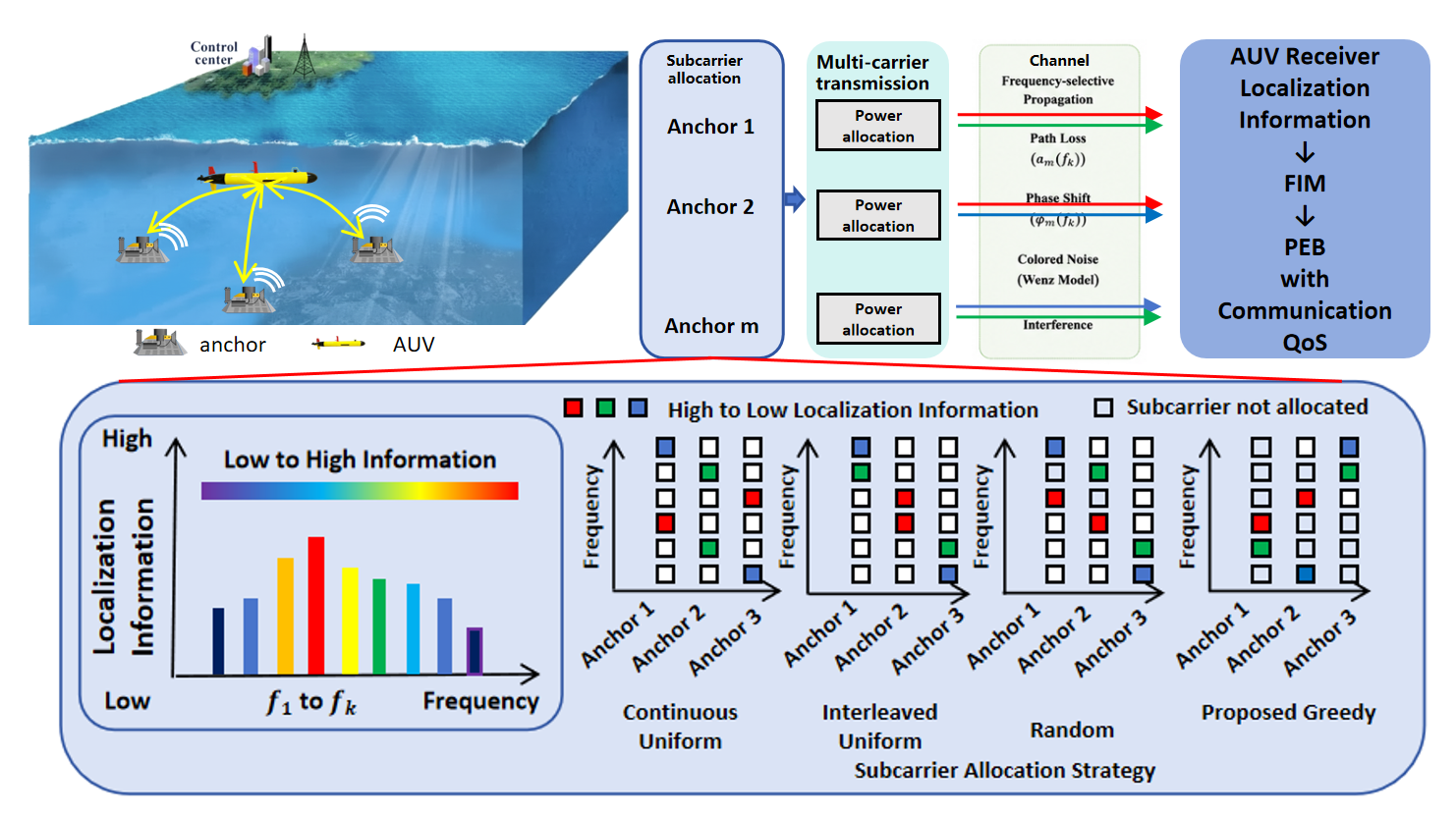}
	\caption{Underwater multi-carrier ISAC system.}\label{fig01}
\end{figure}

This section presents the multi-carrier underwater ISAC system model. It begins with the overall underwater ISAC framework, followed by the multi-carrier signal model, including the transmit signal, underwater acoustic channel, and received signal. Finally, the underwater positioning observation model is established.

\subsection{Multi-carrier Underwater Signal Model}
Consider a multi-carrier underwater ISAC system as shown in Fig.~\ref{fig01}, where an AUV is performing data collection tasks under water, with navigation through anchor nodes. Both AUV and anchor are equipped with single antenna transceivers. Each anchor employs a pilot--data time-division multiplexing (TDM) structure on its allocated subcarriers. Known pilots are transmitted for TOA estimation, whereas communication data are
transmitted during the data interval. Since different anchors occupy disjoint subcarrier sets, their pilot--data switching times need not coincide. 

Let $k \leq K$ denote the index of the $k$th subcarrier with frequency given by $f_k=f_0+(k-1)\Delta f$, where $f_0$ is the carrier frequency and $\Delta f=1/T_s$ is the subcarrier spacing with $T_s$ representing the elementary symbol duration. Let $w_{m,k}\in\left\{0,1\right\}$ denote the subcarrier assignment indicator, where $w_{m,k}=1$ means that the $k$th subcarrier is assigned to the $m$th anchor, and $w_{m,k}=0$ otherwise. To avoid inter-anchor interference, every subcarrier is exclusively assigned to one anchor that 
$\sum_{m=1}^{M}w_{m,k}=1,\quad \forall k.$ In addition, to ensure fairness of communication bandwidth resources, each anchor is allocated the same number of subcarriers, yielding $\sum_{k=1}^{K}w_{m,k}=\frac{K}{M},\quad \forall m.$ 

Considering a frame period $T_f = T_{pl} + T_{da}$, where $T_{pl}$ and $T_{da}$ represent the pilot and data period in a frame, respectively. With power allocation \cite{Sturm2011Waveform,Pucci20225G}, the complex transmit signal of the $m$th anchor during the localization phase with duration $T_{pl}$ is expressed as 
\begin{equation}
	s_m^{pl}(t)=\sum_{k=1}^{K}	\sqrt{z_{m,k}}	w_{m,k}	g_{pl}(t)e^{j2\pi f_k t},\label{eq01}
\end{equation}
where $g_{\mathrm{pl}}(t)$ denotes a known deterministic pilot waveform with effective bandwidth $B_k$, normalized to unit average power over the pilot interval $\frac{1}{T_{\mathrm{pl}}}
\int_0^{T_{\mathrm{pl}}}|g_{\mathrm{pl}}(t)|^2\,dt=1$. The coefficient $z_{m,k}\geq0$ denotes the average transmit power allocated to the $k$th subcarrier of the $m$th anchor.

During the communication phase, the transmitted signal is
\begin{equation}
	s_m^{da}(t)=\sum_{k=1}^{K}\sqrt{z_{m,k}}w_{m,k}x_{m,k}g_{da}(t)e^{j2\pi f_k t},\label{eq02}
\end{equation}
where $x_{m,k}$ denotes the normalized communication symbol satisfying $\mathbb{E}\left\{|x_{m,k}|^2\right\}=1$, and $g_{\mathrm{da}}(t)$ is the pulse-shaping waveform with effective bandwidth $B_k$, normalized such that
$\frac{1}{T_s}\int_0^{T_s}|g_{\mathrm{da}}(t)|^2\,dt=1$. The power allocation satisfies the per-anchor power constraint $\sum_{k=1}^{K}w_{m,k}z_{m,k}\le Z_m,\quad \forall m,$ where $Z_m$ is the maximum transmit power of the $m$th anchor.

The underwater acoustic channel is frequency-selective over the whole system bandwidth but can be regarded as locally flat within each narrowband subcarrier. Therefore, the channel response is modeled as $\rho_{m,k}=a_{m,k}e^{j\phi_{m,k}},$ where $a_{m,k}$ and $\phi_{m,k}$ denote the frequency-dependent amplitude attenuation and phase shift, respectively. The received signal is further influenced by additive colored disturbance consisting of ambient noise and external acoustic interference. The ambient noise follows the Wenz model \cite{Milica2007Wenz}, and the external acoustic interference is approximated as a zero-mean stationary colored Gaussian process following \cite{Snow2009Noise}. Thus, the disturbance is
\begin{equation}
n_m(t)=n_{m,N}(t)+n_{m,I}(t).\label{eq03}
\end{equation}
where $n_{m,N}(t)$ is the ocean background noise, and $n_{m,I}(t)$ refers to fixed external interference caused by factors such as wind power generation and offshore oilfield facilities. Accordingly, the disturbance power spectral density is $N_m(f)=N_{\rm Wenz}(f)+N_{m,I}(f)$, where $N_{\rm Wenz}(f)$ and $N_{m,I}(f)$ are the power spectral density of $n_{m,N}(t)$ and $n_{m,I}(t)$, respectively.

After propagating through the underwater acoustic channel, the multi-carrier signal arrives at the AUV with a propagation delay as $r_m(t)= s_{m}(t)*h_m(t-\tau_m)+ n_m(t)$, where $h_m(t)$ denotes the frequency-selective underwater acoustic channel impulse response, $\tau_m$ is the related transmission delay of signal, and the received multi-carrier signal without noise can be expressed as\footnote{During the localization phase, $x_{m,k}=1$, taking the binary phase shift keying (BPSK) as an example.}
\begin{equation}
	\begin{aligned}
	&\tilde r_m(t-\tau_m) = s_{m}(t)*h_m(t-\tau_m) \\&= \sum_{k=1}^{K} \rho_{m,k} \sqrt{z_{m,k}}x_{m,k}w_{m,k}g\left(t-\tau_m\right)e^{j2\pi f_{k} (t-\tau_m)}.\label{eq4}
	\end{aligned}
\end{equation}
Specifically for localization signal, there is
\begin{equation}
	\begin{aligned}
		\tilde r_m^{pl}(t-\tau_m) = \sum_{k=1}^{K} \rho_{m,k} \sqrt{z_{m,k}}w_{m,k}g_{pl}\left(t-\tau_m\right)e^{j2\pi f_{k} (t-\tau_m)}.\label{eq05}
	\end{aligned}
\end{equation}

\subsection{Communication Rate}
Since each allocated subcarrier is used for both localization pilots and communication data in separate time intervals, the power allocation affects both communication and localization performance. For the $k$th subcarrier of the $m$th anchor, the received per-subcarrier communication Signal-to-Noise Ratio (SNR) is given by
\begin{equation}
\mathrm{SNR}_{m,k}^{\mathrm{com}} = \frac{z_{m,k}|\rho_{m,k}|^2}{	B_k\left(N_{\mathrm{Wenz}}(f_k)+N_{m,I}(f_k)\right)}. \label{eq06}
\end{equation}
Then the achievable communication rate of the $m$th anchor is
\begin{equation}
R_m=\sum_{k=1}^{K}w_{m,k}B_k\log_2\left(1+\mathrm{SNR}_{m,k}^{\mathrm{com}}
\right).\label{eq07}
\end{equation}
This formula indicates that the power $z_{m,k}$ and the subcarrier assignment variable $w_{m,k}$ jointly determine the achievable data rate. In particular, allocating more power to a subcarrier improves its communication SNR, while assigning a subcarrier to a specific anchor determines which anchor can exploit the corresponding frequency-dependent channel condition. Therefore, communication-oriented allocation tends to favor subcarriers with high channel gain and low noise-plus-interference power. However, such allocation may not always coincide with localization-oriented allocation, because localization performance also depends on the delay-resolution term and the geometry-dependent Fisher information contribution. This difference leads to a fundamental tradeoff between communication rate and positioning accuracy in underwater multi-carrier ISAC systems. For localization-oriented resource allocation with communication guarantees, the achievable rate should satisfy a per-anchor rate constraint $R_m\ge R_m^{\min}$.

\subsection{TOA and Localization Model}
The AUV first estimates the propagation delay from the received multi-carrier signals, but due to noise, interference, and frequency-selective channel attenuation, the estimated delay contains uncertainty. Acoustic ray tracing is then used to convert the estimated delay into propagation distance based on the SSP obtained by recent inversion technologies \cite{Huang2024SSP,Huang2025SSP}. Therefore, the final positioning accuracy is
jointly determined by the TOA estimation accuracy and the anchor geometry. 
\subsubsection{TOA Measurement}
The objective of TOA measurement is to estimate the unknown parameter $\tau_m$ from the received signal $r_m(t)$, which subsequently serves as the fundamental observation for underwater positioning. The propagation delay is determined by the actual acoustic propagation path rather than the Euclidean distance because underwater acoustic rays are refracted by the non-uniform SSP. Therefore, the measured TOA can be expressed as
$\tau_m=\int_{\Gamma_m} \frac{1}{c(z)}d\Gamma_m,$ where $\Gamma_m$ denotes the curved propagation path with distance $d_m = \int_{\Gamma_m} d\Gamma_m$ and $c(z)$ is the depth-variant sound speed. Specifically, given the estimated or known SSP, anchor positions, and the AUV’s depth, we can search the grazing angle of signal by matching technologies \cite{Huang2025Optimal}, so as to calculate the acoustic propagation path $d_m$ by ray tracing. Collecting the measurements from all anchors yields the TOA observation vector $\boldsymbol{\tau} =[\tau_1,\tau_2,\cdots,\tau_M]^T$. The estimation accuracy of the target position therefore depends on the estimation accuracy of individual TOA measurements.

\subsubsection{TOA-to-Range Transformation}
Following TOA estimation, the one-way propagation distance is given by $d_m=c_{\mathrm{eff},m}\tau_m.$ where $c_{\mathrm{eff},m}$ is the equivalent sound speed. Assuming a known layered SSP \cite{Huang2024SSP,Huang2025SSP,Wu2026SSP}, the acoustic propagation path is represented by a sequence of ray segments. Accordingly, the propagation distance can be expressed by ray tracing \cite{Huang2025Optimal}: 
\begin{equation}
d_m=\sum_{v=1}^{V}
\frac{\Delta h_v c_0}
{\sqrt{c_0^2-c_{v-1}^2\cos^2\theta_{m,0}}}, \label{eq9}
\end{equation}
where $c_0$ is the initial sound speed, $c_{v-1}$ is the sound speed at the $(v-1)$th layer, $\Delta h_v$ is the depth interval, and $\theta_{m,0}$ denotes the initial grazing angle.


\subsubsection{Geometric Position Model}

Let $\mathbf p\in\mathbb R^D$ denote the unknown true AUV position and let $\mathbf p_m$ denote the known position of the $m$th anchor. For a candidate AUV position $\mathbf p$, the unit direction vector from the $m$th anchor to the AUV is defined as
\begin{equation}
	\bar{\mathbf u}_m(\mathbf p)
	=
	\frac{\mathbf p-\mathbf p_m}
	{\|\mathbf p-\mathbf p_m\|_2}.
	\label{eq:position_direction}
\end{equation}

The true AUV position is unavailable during resource allocation. Instead, a nominal position estimate $\hat{\mathbf p}_0$, obtained from the previous localization result, dead reckoning, or trajectory prediction, is available. Its uncertainty is modeled as
\begin{equation}
	\mathbf p
	=
	\hat{\mathbf p}_0+\boldsymbol{\varepsilon}_p,
	\qquad
	\boldsymbol{\varepsilon}_p
	\sim
	\mathcal N(\mathbf 0,\sigma_p^2\mathbf I_D),
	\label{eq12}
\end{equation}
where $\sigma_p$ is an adjustable position-uncertainty
parameter. Therefore,
$\mathbf p\sim\mathcal N
(\hat{\mathbf p}_0,\sigma_p^2\mathbf I_D)$\footnote{Assuming that the depth is obtained by a depth sensor with ignored errors and the desired position is a two-dimensional horizontal position.}.

\section{Localization-oriented Optimization Problem}
\subsection{TOA CRB with Power Allocation}

\subsubsection{Fisher Information Matrix}
Applying the Fourier transform to the received pilot signal yields the frequency-domain observation
$Y_m^{\mathrm{pl}}(f) =S_m^{\mathrm{pl}}(f)H_m(f)+V_m(f)$, where $H_m(f)=\rho_m(f)e^{-j2\pi f\tau_m}$
denotes the channel response including the propagation delay, and $V_m(f)$ is the frequency-domain disturbance.
The real and imaginary components of the proper complex Gaussian disturbance are uncorrelated and have the same
power spectral density $N_m(f)$. Since this PSD varies with frequency, whitening is used to account for the colored disturbance in the subsequent Fisher information analysis. The noise-free mean of the observation is
$\widetilde{S}_{r,m}^{\mathrm{pl}}(f,\boldsymbol{\xi}) =S_m^{\mathrm{pl}}(f)H_m(f)$, which, under the locally flat subcarrier-channel approximation, can be expressed as
\begin{equation}
	\widetilde S_{r,m}^{\mathrm{pl}}(f;\boldsymbol\xi_m)
	=
	\sum_{k=1}^{K}
	\rho_{m,k}\sqrt{z_{m,k}}w_{m,k}
	G(f-f_k)e^{-j2\pi f\tau_m},
\end{equation}
where $\boldsymbol{\xi}_m \triangleq \left[\tau_m, a_{m,1}, ..., a_{m,K}, \phi_{m,1}, ..., \phi_{m,K}\right]$ is a vector collecting all the channel parameters and $G\left(f\right)$ is the Fourier transform of the reference pilot waveform $g_{\mathrm{pl}}(t)$ ($x_{m,k}=1$ is ignored for localization). 

Equation (11) represents the noise-free mean of the frequency-domain observation: $Y_m^{\mathrm{pl}}(f)
=\widetilde{S}_{r,m}^{\mathrm{pl}}(f,\boldsymbol{\xi}_m) +V_m(f)$.
Assuming that the disturbance PSD $N_m(f)$ is known or accurately estimated by cognitive radio technologies \cite{Karisan2011OFDM}, which is positive over the observation band, whitening is performed using $H_{w,m}(f)=1/\sqrt{N_m(f)}$. The whitened observation is $Y_{\mathrm{w},m}^{\mathrm{pl}}(f)=\bar{S}_m^{\mathrm{pl}}(f,\boldsymbol{\xi}_m)+V_{\mathrm{w},m}(f)$, where
$\bar{S}_m^{\mathrm{pl}}=\widetilde{S}_{r,m}^{\mathrm{pl}}/\sqrt{N_m(f)}$ and $V_{\mathrm{w},m}=V_m/\sqrt{N_m(f)}$.

Let $h_{\mathrm{w},m}(t)$ denote the whitening filter with frequency response $H_{\mathrm{w},m}(f)$.
For a candidate parameter vector $\widetilde{\boldsymbol{\xi}}_m$, the ideal whitened noise-free signal is
$\bar{r}_m^{\mathrm{pl}} (t,\widetilde{\boldsymbol{\xi}}_m) = \tilde{r}_{m}^{\mathrm{pl}}
(t-\tilde{\tau}_m)*h_{\mathrm{w},m}(t)$, where $\tilde{\boldsymbol{\rho}}_m$ collects the candidate
channel coefficients. The real received whitened noisy observation is $\bar{x}_m^{\mathrm{pl}}(t)=r_m^{\mathrm{pl}}(t)*h_{\mathrm{w},m}(t)$. 

%
%
%

Following the standard maximum-likelihood estimation framework for colored Gaussian observations \cite{Karisan2011OFDM}, the Fisher information matrix associated with the propagation delay can be expressed as
\begin{equation}
	\begin{aligned}
		&[\mathbf J_m]_{i,j}= \\
		&\left.
		\operatorname{Re}\left\{
		\int_{-\infty}^{\infty}
		\frac{1}{N_m(f)}
		\frac{\partial
			\widetilde S_{r,m}^{\mathrm{pl}*}
			(f;\widetilde{\boldsymbol\xi}_m)}
		{\partial\widetilde\xi_{m,i}}
		\frac{\partial
			\widetilde S_{r,m}^{\mathrm{pl}}
			(f;\widetilde{\boldsymbol\xi}_m)}
		{\partial\widetilde\xi_{m,j}}
		\,df
		\right\}
		\right|_{\widetilde{\boldsymbol\xi}_m=\boldsymbol\xi_m},
	\end{aligned}
\end{equation}
where $\tilde{\xi}_{j}$ is the jth element of $\tilde{\boldsymbol{\xi}}$, $\frac{\partial \tilde{S}_{r,m}^{pl}(f,\boldsymbol{\xi})}
{\partial \tilde{\xi}_{j}}$ denotes the partial derivative of $\tilde{\boldsymbol{\xi}}$ with respect to $\tilde{\xi}_j$ computed for $\tilde{\boldsymbol{\xi}} =\boldsymbol{\xi} $. 

According to Parseval's theorem, the Fisher information associated with the propagation delay can be simplified into the following closed-form expression
\begin{equation}
	\begin{aligned}
		J_{\tau_m\tau_m}
		={}&4\pi^2\operatorname{Re}\Bigg\{
		\sum_{p=1}^{K}\sum_{k=1}^{K}
		\rho_{m,p}^{*}\rho_{m,k}
		\sqrt{z_{m,p}z_{m,k}}\,
		w_{m,p}w_{m,k}\\
		&\qquad\times
		\int_{-\infty}^{\infty}
		\frac{f^2G^{*}(f-f_p)G(f-f_k)}
		{N_m(f)}\,df
		\Bigg\}.
	\end{aligned}\label{eq13}
\end{equation}

If we define
\begin{equation}
	y_{p,k}^{(2)}\triangleq	\int_{-\infty}^{\infty}	f^2	N_m^{-1}(f)	G^{*}(f-f_p)G(f-f_k)df,
\end{equation}
then we have
\begin{equation}
	J_{\tau_m\tau_m}
	=
	4\pi^2\operatorname{Re}\left\{
	\sum_{p=1}^{K}\sum_{k=1}^{K}
	\rho_{m,p}^{*}\rho_{m,k}
	\sqrt{z_{m,p}z_{m,k}}\,
	w_{m,p}w_{m,k}\,
	y_{p,k}^{(2)}
	\right\}.\label{eq15}
\end{equation}
For orthogonal subcarriers, there is $y_{p,k}^{(2)}=0, p \neq k$, and
\begin{equation}
	J_{\tau_m\tau_m}=	4\pi^2	\sum_{k=1}^{K}	z_{m,k}|\rho_{m,k}|^2	w_{m,k}^{2}\eta_k^{(2)},\label{eq16}
\end{equation}
where $\eta_k^{(2)}=\int_{-\infty}^{\infty}\frac{f^2|G(f-f_k)|^2}{N_m(f)}df$. The obtained Fisher information explicitly shows that the TOA estimation accuracy depends on the combined contribution of all allocated subcarriers, whose contributions are weighted by the channel response, disturbance spectrum, and signal spectrum. 

\subsubsection{Closed-form TOA CRB}
According to the Cramér–Rao theorem, the minimum achievable time delay estimation variance can be expressed as
\begin{equation}
	\operatorname{Var}(\hat{\tau})
	\geq
	\left[\mathbf{J}^{-1}\right]_{1,1}
	=
	\mathrm{CRB}_\tau.
\end{equation}

When the channel amplitudes and phases are known, the conditional TOA CRB is the reciprocal of $J_{\tau_m\tau_m}$
in \eqref{eq16}. Since $w_{m,k}^2=w_{m,k}$, this information can be decomposed into individual subcarrier contributions. Using the locally constant disturbance PSD and the flat-spectrum approximation detailed below, we define the following subcarrier-level information metric.

\textbf{Definition 1:}
For a candidate AUV position $\mathbf p$, the localization
information contributed by the $k$th subcarrier assigned to
the $m$th anchor is defined as
\begin{equation}
	\lambda_{m,k}^{\mathrm{uw}}
	(\mathbf p,z_{m,k})
	=
	\frac{
		4\pi^2 z_{m,k}E_g
		|\rho_{m,k}(\mathbf p)|^2
		\left(
		\frac{B_k^2}{12}+f_k^2
		\right)
	}{
		N_{\mathrm{Wenz}}(f_k)+N_{m,I}(f_k)
	}.
	\label{eq28}
\end{equation}

\textbf{Definition 2:} The accumulated TOA information provided by the subcarriers
assigned to the $m$th anchor is
\begin{equation}
	\Lambda_m
	(\mathbf p;\mathbf W,\mathbf Z)
	=
	\sum_{k=1}^{K}
	w_{m,k}
	\lambda_{m,k}^{\mathrm{uw}}
	(\mathbf p,z_{m,k}).
	\label{eq:accumulated_position_information}
\end{equation}
Accordingly, the conditional TOA CRB for the $m$th anchor is
\begin{equation}
	\mathrm{CRB}_{m,\tau_m}(\mathbf p)
	=
	\frac{1}{
		\Lambda_m(\mathbf p;\mathbf W,\mathbf Z)
	}.
	\label{eq:conditional_toa_crb}
\end{equation}

The quantity $\Lambda_m$ represents the total TOA information collected from all subcarriers assigned to the $m$th anchor.
Therefore, the positioning accuracy is jointly determined by the accumulated ranging information of all anchors and their geometric configuration.

Let $d_m(\mathbf p)$ denote the corresponding acoustic propagation distance between the $m$th anchor and the AUV. The frequency-dependent channel power gain is expressed as
\begin{equation}
	|\rho_{m,k}(\mathbf p)|^2
	=
	10^{
		-\frac{TL(f_k,d_m(\mathbf p))}{10}
	}.
	\label{eq:position_dependent_channel}
\end{equation}
where the transmission loss can be expressed as
\begin{equation}
	TL(f_k,d_m(\mathbf p)) = 10\gamma log_{10}\left(\frac{d_{m}(\mathbf p)}{d_0}\right)+ \alpha(f_k)d_{m,km}(\mathbf p),
\end{equation}
with $\gamma$ describing the signal extension factor. $d_{m,km}$ is the real propagation distance in km, $d_0=1$m is the reference distance, and $\alpha(f_k)$ is the frequency dependent signal absorption coefficient in dB/km. For Thorp' empirical formula, it can be expressed as
\begin{equation}
	\alpha(f)
	=
	\frac{0.11f^{2}}{1+f^{2}}
	+
	\frac{44f^{2}}{4100+f^{2}}
	+
	2.75\times10^{-4}f^{2}
	+
	0.003,
\end{equation}
where $f$ is in kHz. 

Considering the $m$th anchor, there is
\begin{equation}
	|\rho_{m,k}(\mathbf p)|^2
	=
	\left(
	\frac{d_m(\mathbf p)}{d_0}
	\right)^{-\gamma}
	10^{
		-\frac{
			\alpha(f_k)d_{m,\mathrm{km}}(\mathbf p)
		}{10}
	}.
	\label{eq:expanded_channel_gain}
\end{equation}
therefore
\begin{equation}
	\begin{aligned}
		&\lambda_{m,k}^{\mathrm{uw}}
		(\mathbf p,z_{m,k})\\
		&=
		\frac{
			4\pi^2 z_{m,k}E_g
			\left(
			\frac{d_m(\mathbf p)}{d_0}
			\right)^{-\gamma}
			10^{
				-\frac{
					\alpha(f_k)d_{m,\mathrm{km}}(\mathbf p)
				}{10}
			}
			\left(
			\frac{B_k^2}{12}+f_k^2
			\right)
		}{
			N_{\mathrm{Wenz}}(f_k)+N_{m,I}(f_k)
		}.
		\label{eq37}
	\end{aligned}
\end{equation}

The conditional TOA CRB explicitly shows that the achievable ranging accuracy depends on the candidate AUV position through the propagation distance and channel attenuation. Therefore, the position uncertainty should be incorporated when the localization information is used for resource allocation. 
\subsection{Position FIM, CRB and Metric}
The TOA CRB characterizes the achievable ranging accuracy, but it does not explicitly quantify the positioning contribution of individual subcarriers. To facilitate frequency allocation, the following localization information metric is proposed to establish a direct relationship between subcarrier allocation and localization performance.

As assumed previously that the $N_m(f)$ is slowly-varying:
\begin{equation}
	\begin{aligned}
		\eta_k^{(i)}
		&\cong
		\frac{1}{N_m(f_k)}
		\int_{-\infty}^{\infty}
		f^i \left|G(f-f_k)\right|^2 \, df \\
		&=
		\frac{1}{N_m(f_k)}
		\int_{-\infty}^{\infty}
		(f+f_k)^i \left|G(f)\right|^2 \, df .
	\end{aligned}\label{eq27}
\end{equation}
Let $E_g=\int_{-\infty}^{\infty}|G(f)|^2\,df$ denote the pilot waveform energy, and the normalized
spectral moments $\beta_c=E_g^{-1}\int_{-\infty}^{\infty}f^c|G(f)|^2\,df$, $c=1,2$. We can obtain
\begin{equation}
	\eta_k^{(2)}
	=
	\frac{E_g}{N_m(f_k)}
	\left(
	\beta_2 + 2f_k\beta_1 + f_k^2
	\right)
\end{equation}
For analytical tractability, the energy spectrum of the reference pilot is approximated as flat over its effective bandwidth:
\begin{equation}
	|G(f)|^2 \approx
	\begin{cases}
		\dfrac{E_g}{B_k}, & |f|\leq \dfrac{B_k}{2},\\
		0, & \text{otherwise}.
	\end{cases}
\end{equation}
Under the unit-average-power normalization of the reference pilot, Parseval's identity gives
\begin{equation}
	E_g
	=
	\int_{-\infty}^{\infty}|G(f)|^2\,df
	=
	\int_0^{T_{\mathrm{pl,ref}}}
	|g_{\mathrm{pl}}(t)|^2\,dt
	=
	T_{\mathrm{pl,ref}},\label{eq30}
\end{equation}
thus $\beta_1 = 0$ because $f|G(f)|^2$ is odd, and $\beta_2 = \frac{B_k^2}{12}$.

To evaluate the achievable positioning
accuracy, the ranging information must be projected onto the position domain through the geometric observation model. 

Since $d_m(\mathbf p)=c_{\mathrm{eff},m}(\mathbf p)\tau_m$, the conditional range CRB is
\begin{equation}
	\operatorname{CRB}_{m,d_m}(\mathbf p)
	=
	\left(
	\frac{\partial d_m(\mathbf p)}
	{\partial\tau_m}
	\right)^2
	\operatorname{CRB}_{m,\tau_m}(\mathbf p).
	\label{eq:conditional_range_crb_general}
\end{equation}
where $\frac{\partial d_m(\mathbf p)}{\partial \tau_m} = \sum_{v=1}^{V}\frac{\Delta h_{v}c_0}{\sqrt{c_0^2-c^2_{v-1}\cos^2\theta_{m,0}(\mathbf p)}}/\tau_m$. Therefore, the CRB of distance will be
\begin{equation}
	\operatorname{CRB}_{m,d_m}(\mathbf p)
	=
	\frac{
		c_{\mathrm{eff},m}^{2}(\mathbf p)
	}{
		\Lambda_m(\mathbf p;\mathbf W,\mathbf Z)
	}.
	\label{eq:conditional_range_crb}
\end{equation}

According to the chain rule of estimation theory, for a given candidate AUV position $\mathbf p$, the conditional
measurement FIM contributed by the $m$th anchor is
\begin{equation}
	\mathbf J_m(\mathbf p;\mathbf W,\mathbf Z)
	=
	\frac{
		\Lambda_m(\mathbf p;\mathbf W,\mathbf Z)
	}{
		c_{\mathrm{eff},m}^{2}(\mathbf p)
	}
	\bar{\mathbf u}_m(\mathbf p)
	\bar{\mathbf u}_m^{T}(\mathbf p),
	\label{eq:conditional_anchor_fim}
\end{equation}

Since the TOA measurements obtained from different anchors are statistically independent, the conditional measurement FIM for a given candidate position $\mathbf p$ is
\begin{equation}
	\mathbf J_{\mathrm D}
	(\mathbf p;\mathbf W,\mathbf Z)
	=
	\sum_{m=1}^{M}
	\frac{
		\Lambda_m(\mathbf p;\mathbf W,\mathbf Z)
	}{
		c_{\mathrm{eff},m}^{2}(\mathbf p)
	}
	\bar{\mathbf u}_m(\mathbf p)
	\bar{\mathbf u}_m^{T}(\mathbf p).
	\label{eqJpos}
\end{equation}

For the Gaussian position prior in \eqref{eq12}, the prior Fisher information matrix is $\mathbf J_{\mathrm P}=\sigma_p^{-2}\mathbf I_D$. According to the Bayesian Cramer--Rao theorem, the Bayesian information matrix is
\begin{equation}
	\mathbf J_{\mathrm B}(\mathbf W,\mathbf Z)
	=
	\frac{1}{\sigma_p^2}\mathbf I_D
	+
	\mathbb E_{\mathbf p}
	\left[
	\mathbf J_{\mathrm D}
	(\mathbf p;\mathbf W,\mathbf Z)
	\right],
	\label{eq:bayesian_fim}
\end{equation}
where the expectation is taken with respect to $\mathbf p\sim\mathcal N (\hat{\mathbf p}_0,\sigma_p^2\mathbf I_D)$.

\textbf{Definition 3:} To evaluate the effectiveness of the proposed A-optimal frequency allocation strategy, the Bayesian positioning error bound (BPEB) is defined as
\begin{equation}
	\operatorname{BPEB}(\mathbf W,\mathbf Z)
	=
	\sqrt{
		\operatorname{tr}
		\left[
		\mathbf J_{\mathrm B}^{-1}
		(\mathbf W,\mathbf Z)
		\right]
	}.
	\label{eq40}
\end{equation}
The BPEB lower-bounds the Bayesian root-mean-square position error averaged over both the prior position uncertainty and the measurement observations. The expectation is numerically approximated as $\mathbb E_{\mathbf p}[\mathbf J_{\mathrm D}(\mathbf p)]
\approx N_s^{-1}\sum_{s=1}^{N_s} \mathbf J_{\mathrm D}(\mathbf p^{(s)})$ with $N_s$ positioning samples, where $\mathbf p^{(s)}\sim
\mathcal N(\hat{\mathbf p}_0,\sigma_p^2\mathbf I_D)$.

The derived CRB depends jointly on the localization information accumulated from all anchors and their spatial geometry. Therefore, localization performance can be improved not only by increasing the localization information of individual links but also by allocating subcarrier more effectively across multiple anchors. The formula \eqref{eq40} provides a theoretical standard for evaluating positioning performance, but the root mean square error (RMSE) is still used to measure positioning accuracy in practice, which can be expressed as $E_{rmse} = \|\mathbf{p}-\mathbf{p}_{es}\|_2$ with $\mathbf{p}=[x,y,z]^T$ and $\mathbf{p}_{es}[x_{es},y_{es},z_{es}]^T$ denoting the actual and estimated AUV's position, respectively.

\textbf{Definition 4:} For the position prior distribution \(f_{\mathbf p}(\mathbf p)\), define the prior-averaged information matrices as
\begin{equation}
	\mathbf Q_{m,k}(z_{m,k})
	\triangleq
	\mathbb E_{\mathbf p}
	\left[
	\frac{
		\lambda_{m,k}^{\mathrm{uw}}
		(\mathbf p,z_{m,k})
	}{
		c_{\mathrm{eff},m}^{2}(\mathbf p)
	}
	\bar{\mathbf u}_m(\mathbf p)
	\bar{\mathbf u}_m^{T}(\mathbf p)
	\right].
	\label{eq41}
\end{equation}
The Bayesian information matrix can then be expressed as
\begin{equation}
	\mathbf J_{\mathrm B}(\mathbf W,\mathbf Z)
	=
	\frac{1}{\sigma_p^2}\mathbf I_D
	+
	\sum_{m=1}^{M}\sum_{k=1}^{K}
	w_{m,k}\mathbf Q_{m,k}(z_{m,k}).
	\label{eq:bayesian_fim_decomposition}
\end{equation}
\begin{proposition}\label{pro1} The effects of $\mathbf W$ and $\mathbf Z$ on the Bayesian positioning performance are characterized by the prior-averaged information matrices $\{\mathbf Q_{m,k}(z_{m,k})\}$.
\end{proposition}

\begin{proof}Substituting $\Lambda_m(\mathbf p;\mathbf W,\mathbf Z)=\sum_{k=1}^{K}w_{m,k} \lambda_{m,k}^{\mathrm{uw}}(\mathbf p,z_{m,k})$ into the conditional data FIM gives
\begin{equation}
	\mathbf J_{\mathrm D}
	(\mathbf p;\mathbf W,\mathbf Z)
	=
	\sum_{m=1}^{M}\sum_{k=1}^{K}
	w_{m,k}
	\frac{
		\lambda_{m,k}^{\mathrm{uw}}
		(\mathbf p,z_{m,k})
	}{
		c_{\mathrm{eff},m}^{2}(\mathbf p)
	}
	\bar{\mathbf u}_m(\mathbf p)
	\bar{\mathbf u}_m^{T}(\mathbf p).
\end{equation}
Taking the expectation over the position prior and adding the Gaussian prior information $\sigma_p^{-2}\mathbf I_D$ yields \eqref{eq37}. Therefore, the dependence of the BPEB on $\mathbf W$ and $\mathbf Z$ is characterized by the prior-averaged information matrices $\{\mathbf Q_{m,k}(z_{m,k})\}$. This completes the proof.
\end{proof}

To this end, Proposition \ref{pro1}\footnote{in which the real signal propagation path can be calculated through ray tracing with fixed network topology \cite{Huang2025Optimal}.} shows that each anchor--subcarrier pair contributes a prior-averaged information matrix to the Bayesian FIM. This additive structure motivates a greedy allocation strategy that sequentially selects the pair producing the largest reduction in the Bayesian A-optimal objective.

\subsection{Problem Formulation}
Based on the derived localization performance metric, the localization-priority optimization problem is formulated as
\begin{subequations}
	\begin{align}
		\mathcal P_1^{\mathrm B}:\quad	\min_{\mathbf W,\mathbf Z}\quad
		&
		\operatorname{tr}
		\left[
		\mathbf J_{\mathrm B}^{-1}
		(\mathbf W,\mathbf Z)
		\right]
		\label{problem:bayesian_objective}\\
		\mathrm{s.t.}\quad
		& R_m(\mathbf W,\mathbf Z)\geq R_m^{\mathrm{th}},	\quad \forall m,\\
		& \sum_{m=1}^{M} w_{m,k}=1,\quad \forall k,\\
		& \sum_{k=1}^{K} w_{m,k}=\frac{K}{M},\quad \forall m,\\
		& \sum_{k=1}^{K} w_{m,k}z_{m,k}\le Z_m,\quad \forall m,\\
		& z_{m,k}\ge 0,\quad \forall m,k,\\
		& w_{m,k}\in\left\{0,1\right\},\quad \forall m,k.
	\end{align}
\end{subequations}
The first constraint guarantees that the communication throughput of each anchor satisfies the minimum QoS requirement $R_{\rm th}$. The second and third constraints enforce orthogonal subcarrier assignment and equal subcarrier allocation among anchors, respectively. The fourth and fifth constraints correspond to the per-anchor power budget and non-negative power allocation. The last constraint specifies the binary nature of the subcarrier assignment variables.

Problem $\mathcal P_1^{\mathrm B}$ minimizes the Bayesian positioning error under communication QoS, orthogonal subcarrier-assignment, and per-anchor power constraints. The objective explicitly accounts for the uncertainty of the nominal AUV position used during resource allocation.

In practical implementation, the nominal position $\hat{\mathbf p}_0$ can be obtained from the previous localization result, dead reckoning, or trajectory prediction. The adjustable parameter $\sigma_p$ characterizes the corresponding position uncertainty. Resource allocation is then performed using the prior distribution $\mathcal N(\hat{\mathbf p}_0,\sigma_p^2\mathbf I_D)$ rather
than treating $\hat{\mathbf p}_0$ as the true AUV position.

\section{Localization-Oriented Joint Resource Allocation}
\subsection{Properties of Localization Information}
\subsubsection{Distance Dependence}
To quantitatively characterize the impact of propagation distance on localization capability, the following proposition establishes the monotonic relationship between propagation distance and the proposed localization information metric.

\begin{proposition}\label{pro2}
	The underwater localization information $\lambda^{uw}_{m,k}$ is a monotonically decreasing function of the propagation distance $d_{m,km}$ that $\frac{\partial\lambda^{uw}_{m,k}}	{\partial d_{m,km}}<0$.
\end{proposition}

\begin{proof}
	According to \eqref{eq37}, $\lambda^{uw}_{m,k}	\propto	d_{m,km}^{-\gamma} 10^{-\frac{\alpha(f_k)d_{m,km}}{10}}$. Taking the logarithm yields
	\begin{equation}
		\ln\lambda^{uw}_{m,k}	=	-\gamma\ln d_{m,km}	-\frac{\ln10}{10}\alpha(f_k)d_{m,km}	+C,
	\end{equation}
	where $C$ is independent of $d_{m,km}$.
	Differentiating both sides gives
	\begin{equation}
		\frac{\partial
			\ln\lambda^{uw}_{m,k}}
		{\partial d_{m,km}}
		=
		-\frac{\gamma}{d_{m,km}}
		-\frac{\ln10}{10}\alpha(f_k)<0.
	\end{equation}
	Since
	\(
	\lambda^{uw}_{m,k}>0
	\),
	we have
	\begin{equation}
		\frac{\partial
			\lambda^{uw}_{m,k}}
		{\partial d_{m,km}}
		=
		\lambda^{uw}_{m,k}
		\frac{\partial
			\ln\lambda^{uw}_{m,k}}
		{\partial d_{m,km}}
		<0.
	\end{equation}
	The proof is completed.
\end{proof}
\begin{remark}
	Proposition \ref{pro2} provides a fundamental ordering principle for localization-oriented resource allocation, indicating that propagation distance is a dominant factor determining the contribution of each subcarrier. This result shows the inherent information imbalance among anchors and serves as a key guideline for designing distance-aware allocation strategies and greedy optimization procedures.
\end{remark}

\subsubsection{Frequency Dependence}
Unlike propagation distance, the influence of carrier frequency on localization performance is more intricate. Increasing the carrier frequency improves the delay estimation sensitivity because the phase variation induced by propagation delay becomes more significant at higher frequencies. However, underwater acoustic absorption also increases rapidly with frequency, resulting in more severe propagation loss and reduced received signal power. Therefore, the localization information is jointly affected by these competing mechanisms. The following proposition characterizes this fundamental tradeoff.

\begin{proposition}\label{pro3}
There exists an optimal carrier frequency $f_k^\star$ that maximizes the localization information $f_k^\star	=	\arg\max_f	\lambda(f,d)$.
\end{proposition}

\begin{proof}
	Taking the natural logarithm of both sides of \eqref{eq37} gives
	\begin{align}
		\ln \lambda^{uw}_{m,k}
		=& \ln\left(4\pi^2z_{m,k}E_g\right)
		-\gamma \ln d_{m,km} -\ln\left(N_{m}(f_k)\right) \notag\\
		&+\ln\left(\frac{B_k^2}{12}+f_k^2\right)
		-\frac{\ln 10}{10}\alpha(f_k)d_{m,km} .
	\end{align}
	
	Differentiating $\ln \lambda^{uw}_{m,k}$ with respect to $f_k$ yields
	\begin{align}
		\frac{\partial \ln \lambda^{uw}_{m,k}}{\partial f_k}
		=&
		\frac{2f_k}
		{\frac{B_k^2}{12}+f_k^2}
		-\frac{\ln 10}{10}d_{m,km}\alpha'(f_k)-\frac{N'_{m}(f_k)}{N_{m}(f_k)} .
	\end{align}
	
	Since $\lambda^{uw}_{m,k}>0$, we have
	\begin{equation}
		\frac{\partial \lambda^{uw}_{m,k}}{\partial f_k}
		=
		\lambda^{uw}_{m,k}
		\frac{\partial \ln \lambda^{uw}_{m,k}}{\partial f_k}.
	\end{equation}
	
	Therefore, the sign of 	$\partial \lambda^{uw}_{m,k}/\partial f_k$ is determined by the balance among the above three terms. The first term is positive because a higher carrier frequency increases the sensitivity of the signal phase to propagation delay, which improves TOA estimation accuracy. The second term is negative because the acoustic absorption coefficient generally increases with frequency \(\alpha'(f_k)>0\). The third term captures the influence of the colored ambient noise and external interference spectrum. Hence, increasing the carrier frequency improves localization information only when the delay-resolution gain dominates the absorption loss and disturbance-spectrum variation. Otherwise, the localization information decreases with frequency. 
\end{proof}

\subsubsection{Frequency Ordering Principle}
The following theorem establishes a distance-dependent frequency ordering property, which reveals how frequency resources should be matched with anchors at different propagation distances.

\begin{theorem}\label{the1}
	\label{thm:multi_anchor_frequency_ordering}
	Consider two anchors $i$ and $j$ with propagation distances
	$d_i<d_j$. the ratio of localization information is monotonically increasing with frequency: $\partial\frac{\lambda_i(f)}{\lambda_j(f)}/\partial f>0.$
\end{theorem}

\begin{proof}
	According to (\ref{eq28}), the localization information	of the $m$th anchor at frequency $f$ can be written as
	\begin{equation}
		\lambda_m(f)=C	d_m^{-\gamma}	10^{-\frac{\alpha(f)d_m}{10}}	\frac{\left(\frac{B^2}{12}+f^2\right)}	{N_m(f)},
	\end{equation}
	where $C$ is independent of $m$ and $f$, and $N_m(f)$ denotes	the disturbance power spectrum.
	
	For any two anchors $i$ and $j$, the ratio is
	\begin{equation}
		R_{i,j}(f)	=\frac{\lambda_i(f)}{\lambda_j(f)}=\left(\frac{d_i}{d_j}\right)^{-\gamma}10^{-\frac{\alpha(f)(d_i-d_j)}{10}} .
	\end{equation}
	Taking the natural logarithm gives
	\begin{equation}
		\ln R_{i,j}(f)	=	-\gamma	\ln	\left(	\frac{d_i}{d_j}	\right)	-	\frac{\ln 10}{10}	\alpha(f)(d_i-d_j).
	\end{equation}
	
	Differentiating with respect to $f$ yields
	\begin{equation}
		\frac{\partial}{\partial f}	\ln R_{i,j}(f)	=-	\frac{\ln 10}{10}	(d_i-d_j)\alpha'(f).
	\end{equation}
	
	Since $d_i<d_j$ and the underwater acoustic absorption coefficient satisfies $\alpha'(f)>0$, we have
	\begin{equation}
		\frac{\partial}{\partial f}	\ln R_{i,j}(f)>0 .
	\end{equation}
	
	Because $R_{i,j}(f)>0$, it follows that
	\begin{equation}
		\frac{\partial R_{i,j}(f)}{\partial f}	=	R_{i,j}(f)	\frac{\partial}{\partial f}	\ln R_{i,j}(f)>0 .
	\end{equation}
	The proof is completed.
\end{proof}

Theorem \ref{the1} indicates that higher-frequency subcarriers provide a larger relative localization benefit to nearer anchors, whereas lower-frequency subcarriers are relatively more favorable for distant anchors.

\subsection{Greedy Subcarrier Allocation with Local Power Refinement}
The optimization problem $P_1^{\mathrm B}$ jointly optimizes the binary subcarrier allocation matrix $\mathbf W$ and the continuous power allocation matrix $\mathbf Z$, which leads to a mixed-integer nonlinear programming problem. An exhaustive search can provide the optimal solution for small-scale systems; however, its computational complexity grows exponentially with the number of anchors and subcarriers.

Motivated by the Bayesian FIM decomposition in Proposition~\ref{pro1}, we develop a greedy allocation algorithm based on the prior-averaged information matrices $\mathbf Q_{m,k}$. A reference equal-power allocation is first used during the subcarrier-assignment stage. At each iteration, the feasible anchor--subcarrier pair producing the largest reduction in the Bayesian A-optimal objective is selected. After obtaining the subcarrier assignment, the power coefficients are locally refined under the communication and power constraints.

\begin{algorithm}[t]
	\caption{Bayesian Localization-Oriented Greedy Allocation}
	\label{alg:bayesian_allocation}
	\begin{algorithmic}[1]
		\STATE Draw $N_s$ position samples from
		$\mathcal N(\hat{\mathbf p}_0,\sigma_p^2\mathbf I_D)$
		\STATE Set $z_m^{(0)}=MZ_m/K$ and compute
		$\mathbf Q_{m,k}^{(0)}=\mathbf Q_{m,k}(z_m^{(0)})$
		\STATE Initialize
		$\mathbf W=\mathbf 0$, $n_m=0$, and
		$\mathbf J=\sigma_p^{-2}\mathbf I_D$
		\FOR{$t=1,\ldots,K$}
		\STATE Evaluate all feasible $(m,k)$ satisfying
		$\sum_{i=1}^{M}w_{i,k}=0$ and $n_m<K/M$
		\STATE Compute $
		\Delta_{m,k}
		=
		\operatorname{tr}(\mathbf J^{-1})
		-
		\operatorname{tr}
		[(\mathbf J+\mathbf Q_{m,k}^{(0)})^{-1}]$
		\STATE Select
		$(m^\star,k^\star)
		=\arg\max_{m,k}\Delta_{m,k}$
		\STATE Set $w_{m^\star,k^\star}=1$ and update
		$n_{m^\star}$ and
		$\mathbf J\leftarrow
		\mathbf J+\mathbf Q_{m^\star,k^\star}^{(0)}$
		\ENDFOR
		\STATE Refine $\mathbf Z$ with fixed $\mathbf W$ by minimizing
		$\operatorname{tr}
		[\mathbf J_{\mathrm B}^{-1}(\mathbf W,\mathbf Z)]$
		under the communication and power constraints
		\STATE Output $\mathbf W^\star$ and $\mathbf Z^\star$
	\end{algorithmic}
\end{algorithm}

\subsection{Computational Complexity Analysis}

The prior-averaged information matrices $\mathbf Q_{m,k}^{(0)}$ are computed once before the greedy allocation, resulting in complexity $O(N_sMKD^2)$, where $N_s$ denotes the number of position samples. The greedy stage evaluates at most $M(K-t+1)$ candidate pairs at the
$t$th iteration. Since each candidate requires inversion of a $D\times D$ information matrix, its total complexity is $O(MK^2D^3)$. The local power refinement with $N_p$ iterations requires $O[N_p(KD^2+D^3)]$. Thus, the overall complexity is
\begin{equation}
	O\left( N_sMKD^2+MK^2D^3+N_p(KD^2+D^3)\right).
\end{equation}
Since $D$ and $N_s$ are small and fixed, this reduces to $O(MK^2+N_pK)$, which has the same polynomial order as the nominal-position implementation.

\section{Experimental Result and Analysis}
\subsection{Simulation Setup}
To evaluate the effectiveness of the proposed localization-oriented frequency allocation strategy, simulations are conducted under a multi-carrier underwater ISAC system. Unless otherwise specified, the simulation parameters listed in Table~\ref{tab01} are adopted throughout this section.


To evaluate the effectiveness of the proposed A-optimal frequency allocation strategy, several representative allocation methods are compared, including continuous uniform allocation (Uniform-C), interleaved uniform allocation (Uniform-I), random allocation, and high-frequency-first allocation. The primary localization metric is the BPEB, and the effective communication throughput is adopted to evaluate the communication performance under different localization durations. Generally, each simulation result is obtained by averaging over 100 independent Monte Carlo trials.
\begin{table}[!htbp]
	\caption{Simulation Parameters}
	\label{tab01}
	\centering
	\begin{tabular}{cc}
		\hline
		Parameter & Value\\
		\hline
		Ocean depth & 2900 m\\
		Depth of AUV & 10 m\\
		Anchor topology & Equilateral triangle \\
		Default AUV position & (0,0,-10) \\
		Rate constraint & variable\\
		Sampling frequency & 100 kHz\\
		Frame duration & 1 s, 20 s(Fig.9) \\
		Number of anchors $M$ & 3,4,5\\
		Number of subcarriers $K$ & 6,9,12;8;10\\
		System bandwidth $B_{\rm sys}$ & 12,16,18,20 kHz \\
		Carrier frequency range & 8--20 (up to 28) kHz \\
		Subcarrier bandwidth $B_k$ & 2 kHz \\
		Spreading factor $\gamma$ & 1.5 \\
		Propagation model & Thorp model \\
		Noise model & Wenz model (shipping 0, wind speed 4 m/s) \\
		External interference & Not considered\\
		Transmit Level & 170 dB \\
		Position samples $N_s$ & 20 \\
		Position-uncertainty $\sigma_p$ & 1 m \\
		\hline
	\end{tabular}
\end{table}

\subsection{Validation of Localization Information Properties}
\begin{figure*}[!htbp]
	\centering
	\subfloat[$\lambda^{uw}_{m,k}$ versus frequency.]{\includegraphics[width=0.3\textwidth]{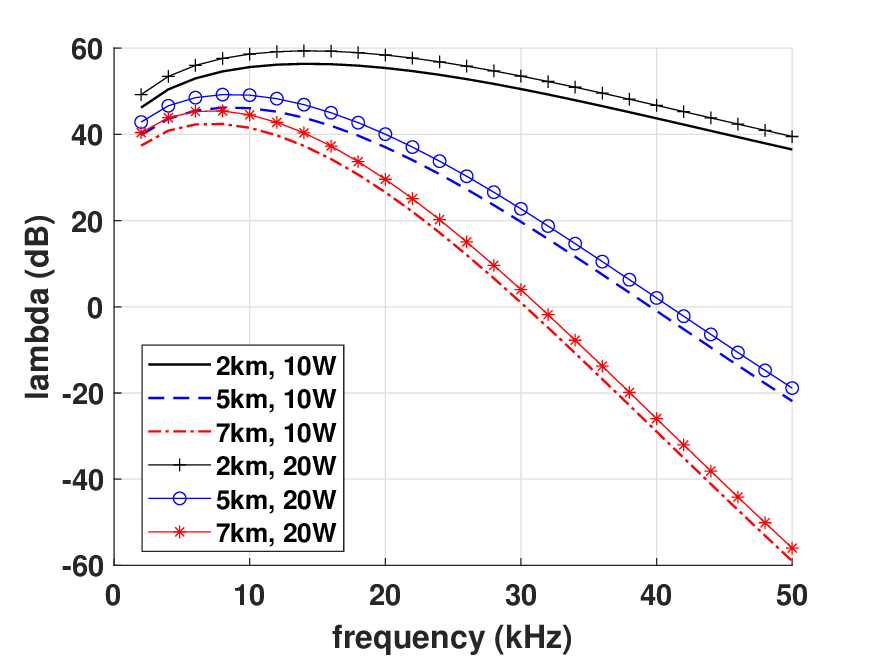}\label{fig03A}}
	\subfloat[$\lambda^{uw}_{m,k}$ versus distance.]{\includegraphics[width=0.3\textwidth]{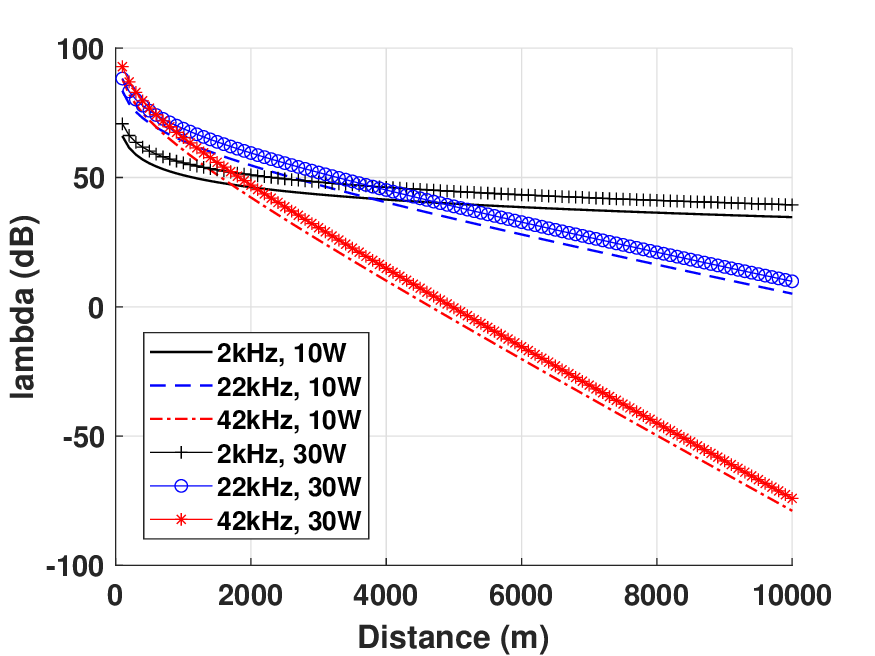}\label{fig03B}}
	\subfloat[$\lambda^{uw}_{m,k}$ versus power.]{\includegraphics[width=0.3\textwidth]{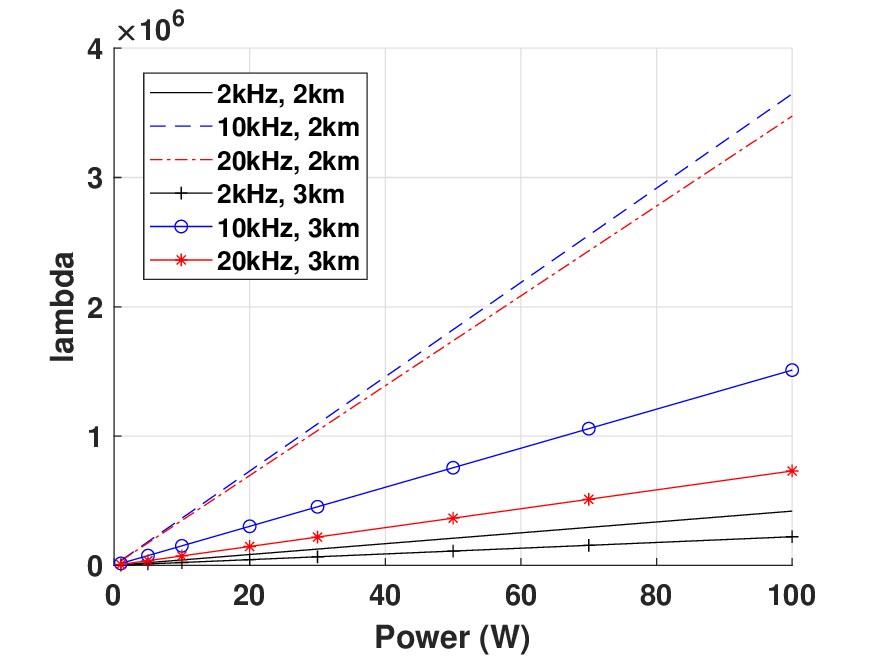}\label{fig03C}}
	\caption{Influence of frequency, distance and transmit power on localization information $\lambda^{uw}_{m,k}$.}\label{fig03}
\end{figure*}
This subsection validates the theoretical properties of the proposed localization information metric. In particular, the impacts of carrier frequency, propagation distance, and transmit power on the localization information are investigated to verify the analytical results derived in Section \uppercase\expandafter{\romannumeral4}.

\subsubsection{Frequency Effect}
Fig.~\ref{fig03A} shows the effect of carrier frequency under different propagation distances. The localization information first increases with frequency because higher frequencies provide better delay resolution, but then decreases when absorption loss becomes dominant. In addition, the optimal frequency shifts to lower values as the distance increases. These results confirm the tradeoff between ranging resolution and frequency-dependent attenuation, showing that simply using higher-frequency subcarriers does not always improve localization performance.

\subsubsection{Distance Effect}
To verify Proposition \ref{pro2}, Fig.~\ref{fig03B} shows the localization information versus propagation distance for different carrier frequencies. The localization information decreases with distance for all frequencies due to geometric spreading and absorption loss. Higher-frequency subcarriers degrade faster because of stronger acoustic attenuation, indicating that their localization advantage becomes weaker at long ranges.

\subsubsection{Power Effect}
Fig.~\ref{fig03C} shows the localization information coefficient $\lambda^{uw}_{m,k}$ versus transmit power under different propagation distances and carrier frequencies. $\lambda^{uw}_{m,k}$ increases almost linearly with transmit power, confirming that higher power directly improves localization information. The result indicates that power allocation should jointly account for propagation distance and carrier frequency to improve localization performance.

\subsection{Performance of Joint Resource Allocation}
\subsubsection{Prior-Averaged Information Distribution}
\begin{figure}[!htbp]
	\centering
	\subfloat[]{\includegraphics[width=0.24\textwidth]{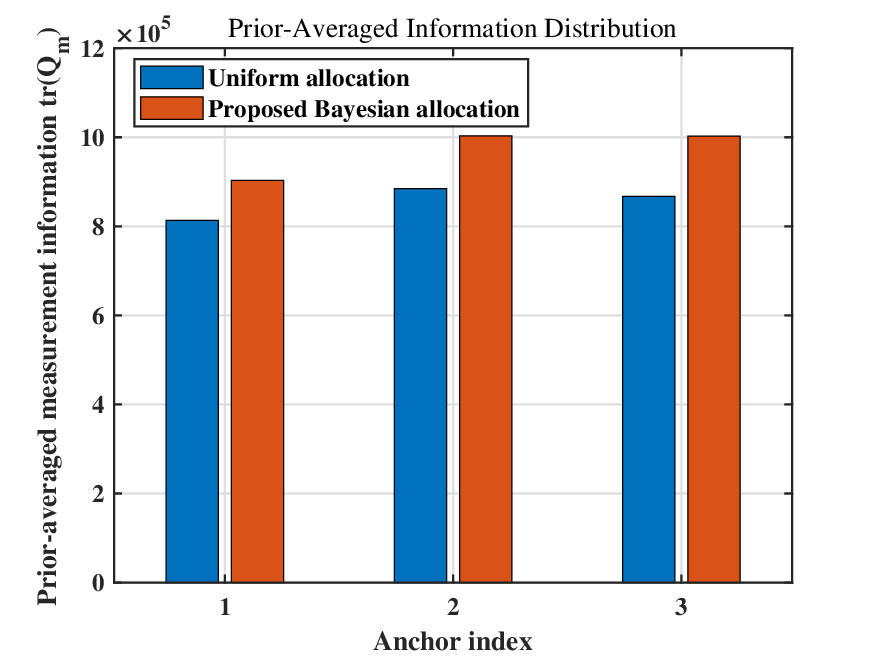}\label{fig04a}}
	\subfloat[]{\includegraphics[width=0.24\textwidth]{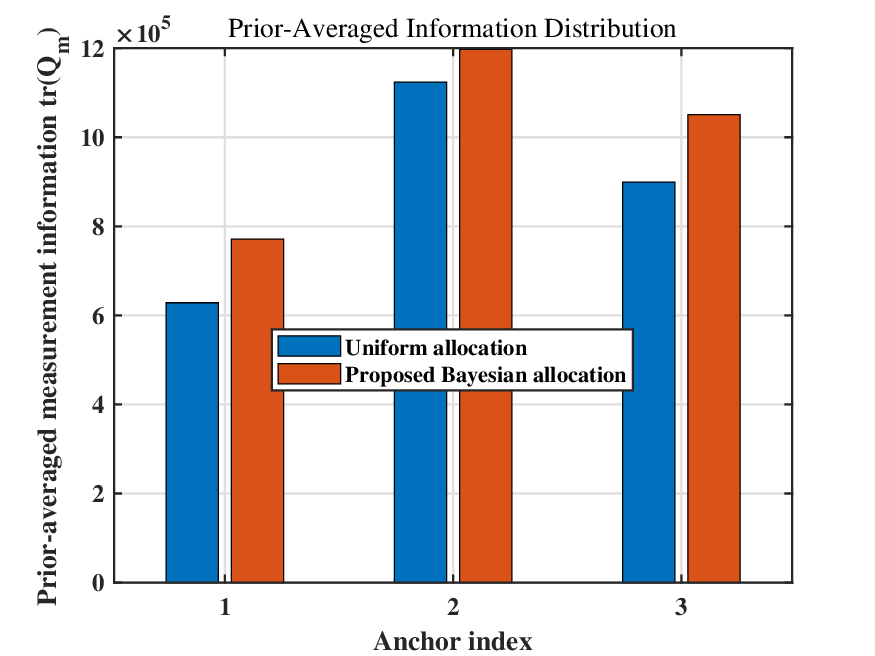}\label{fig04b}}
	\caption{Prior-averaged measurement-information distributions under (a) symmetric and (b) asymmetric three-anchor geometries.}\label{fig04}
\end{figure}
\begin{figure*}[!htbp]
	\centering
	\subfloat[]{\includegraphics[width=0.2\textwidth]{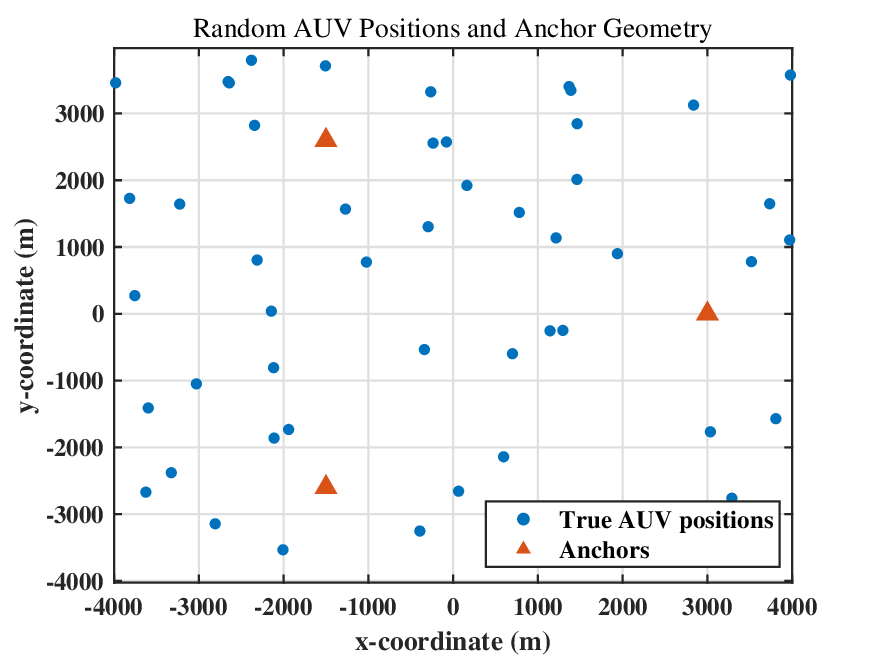}\label{fig05a}}
	\subfloat[]{\includegraphics[width=0.2\textwidth]{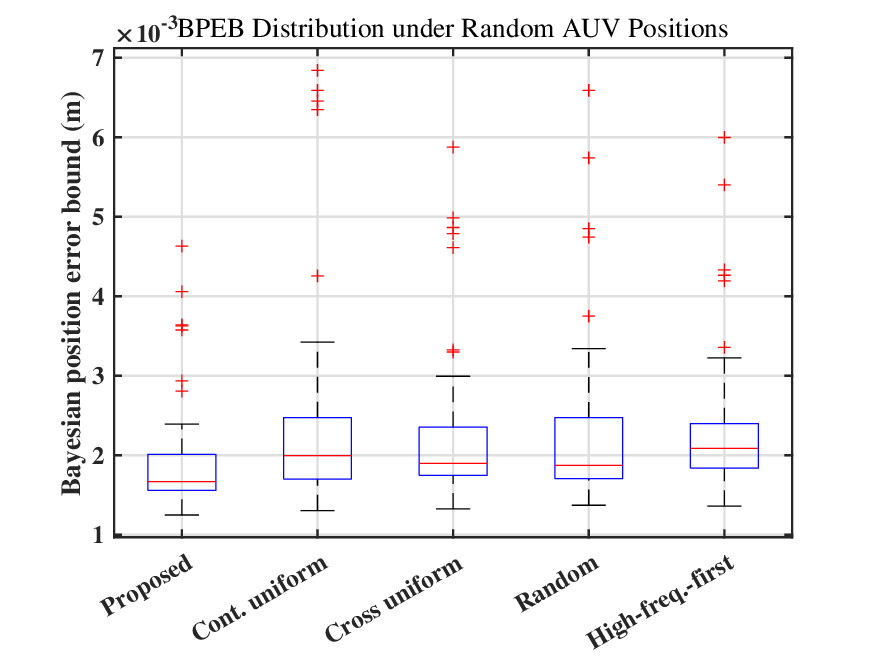}\label{fig05b}}
	\subfloat[]{\includegraphics[width=0.2\textwidth]{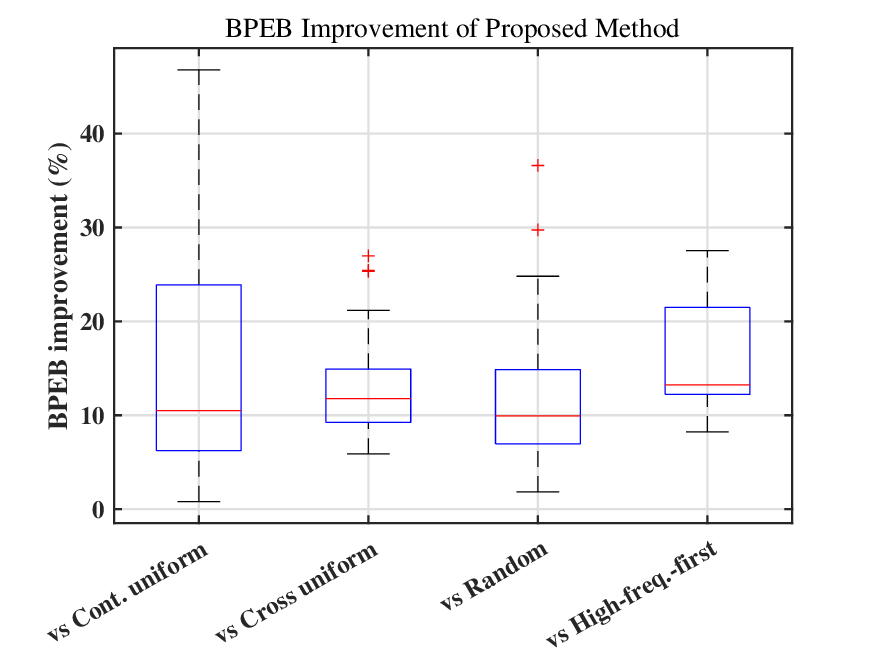}\label{fig05c}}
	\subfloat[]{\includegraphics[width=0.2\textwidth]{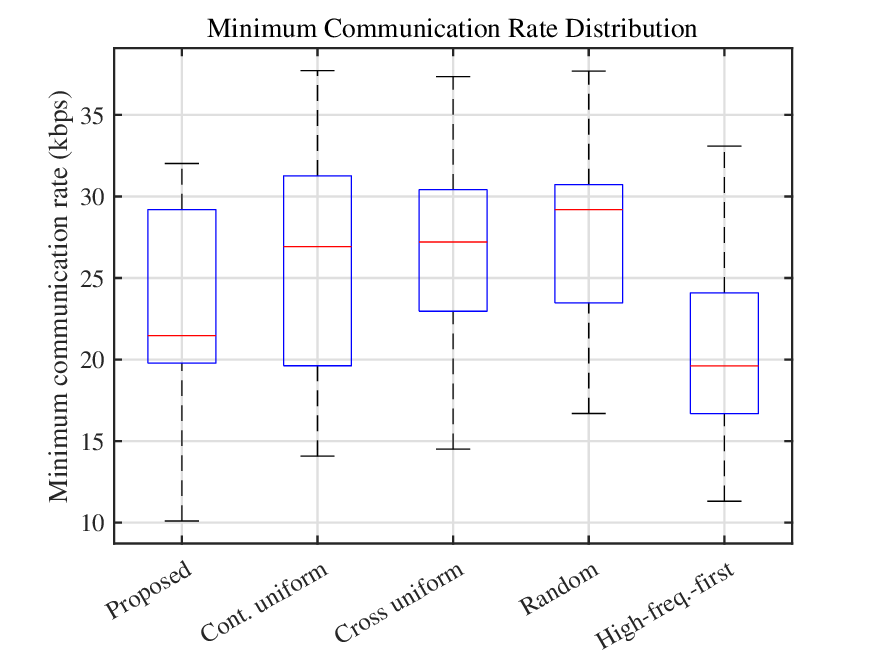}\label{fig05d}}\\
	\subfloat[]{\includegraphics[width=0.2\textwidth]{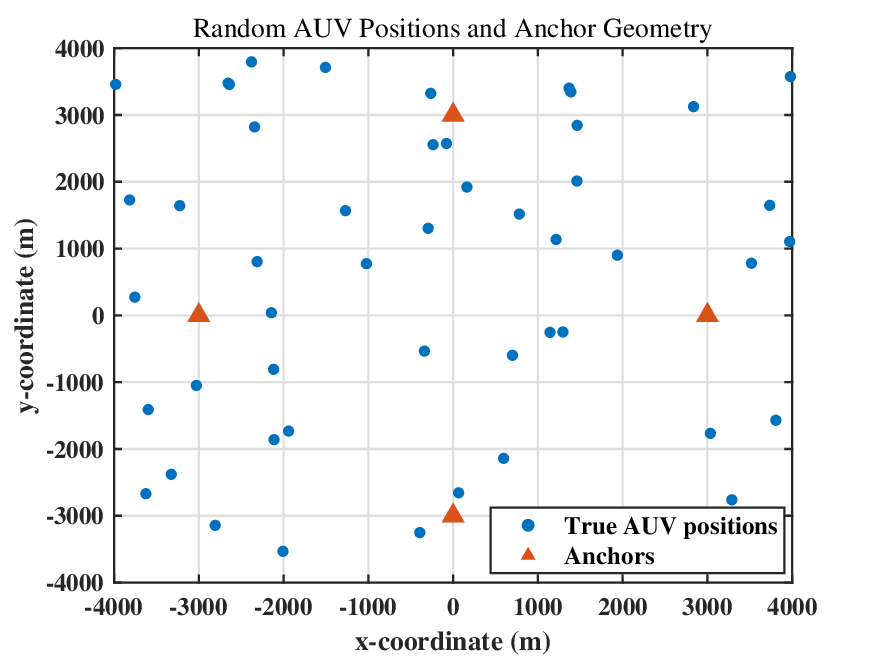}\label{fig05e}}
	\subfloat[]{\includegraphics[width=0.2\textwidth]{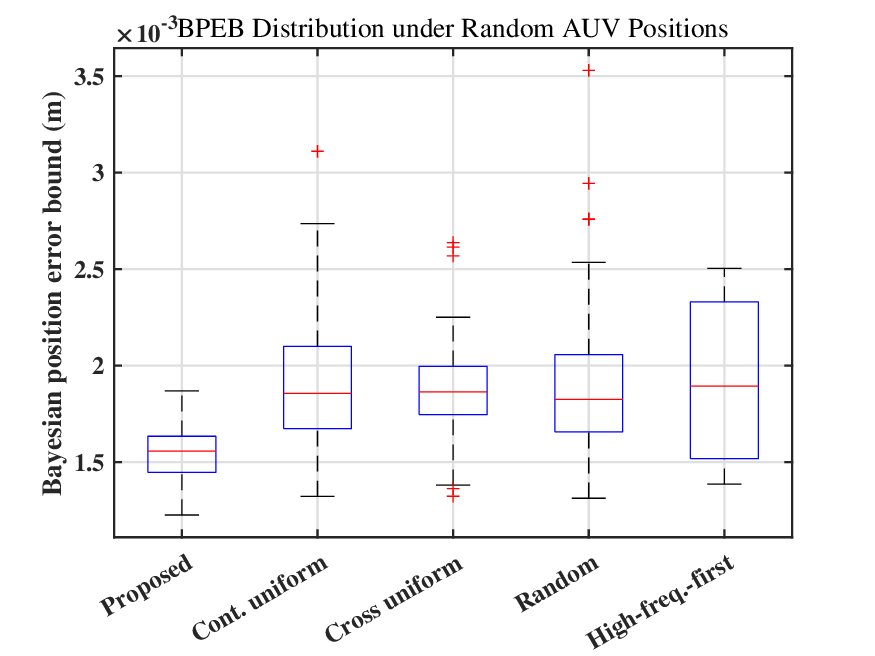}\label{fig05f}}
	\subfloat[]{\includegraphics[width=0.2\textwidth]{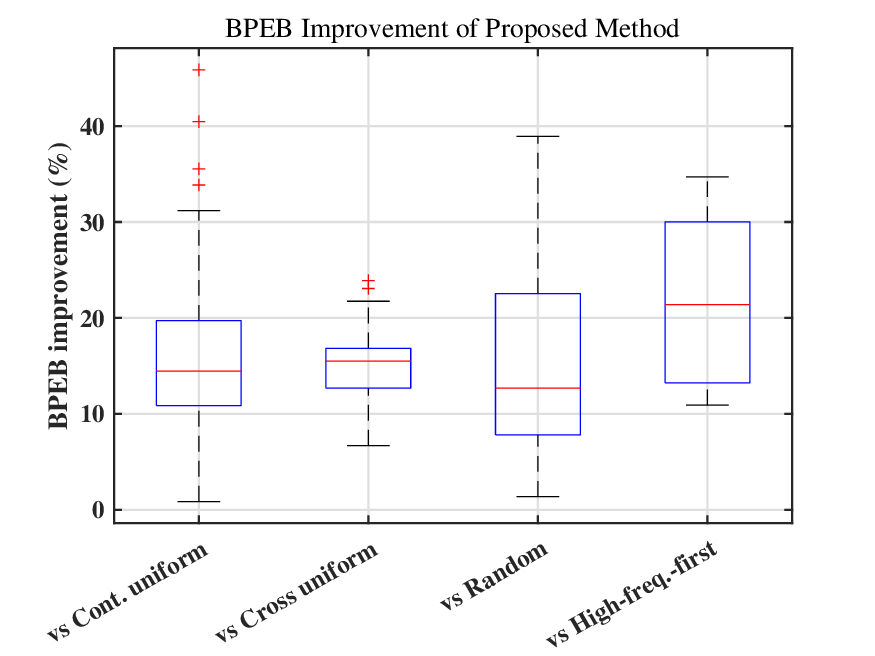}\label{fig05g}}
	\subfloat[]{\includegraphics[width=0.2\textwidth]{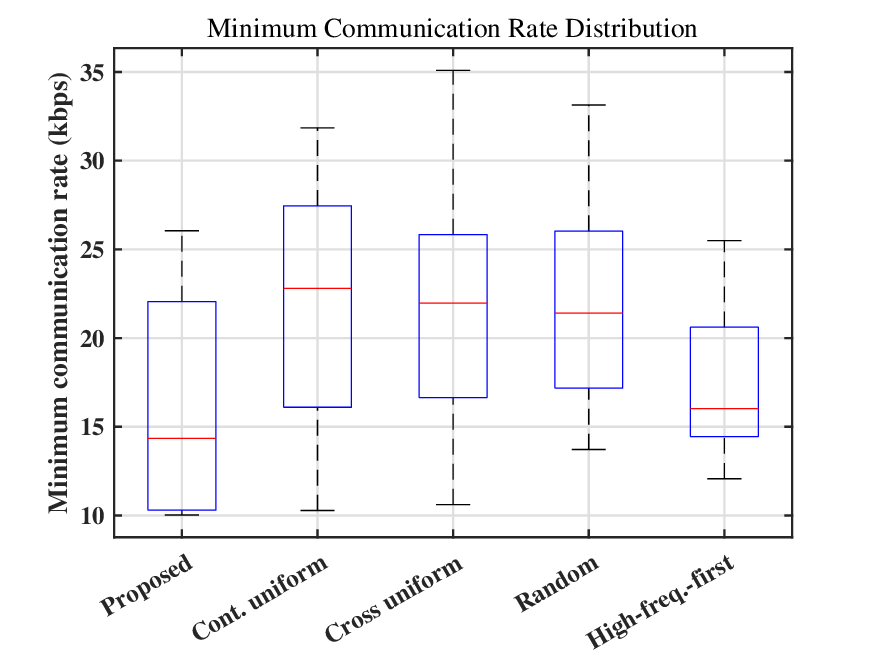}\label{fig05h}}\\
	\subfloat[]{\includegraphics[width=0.2\textwidth]{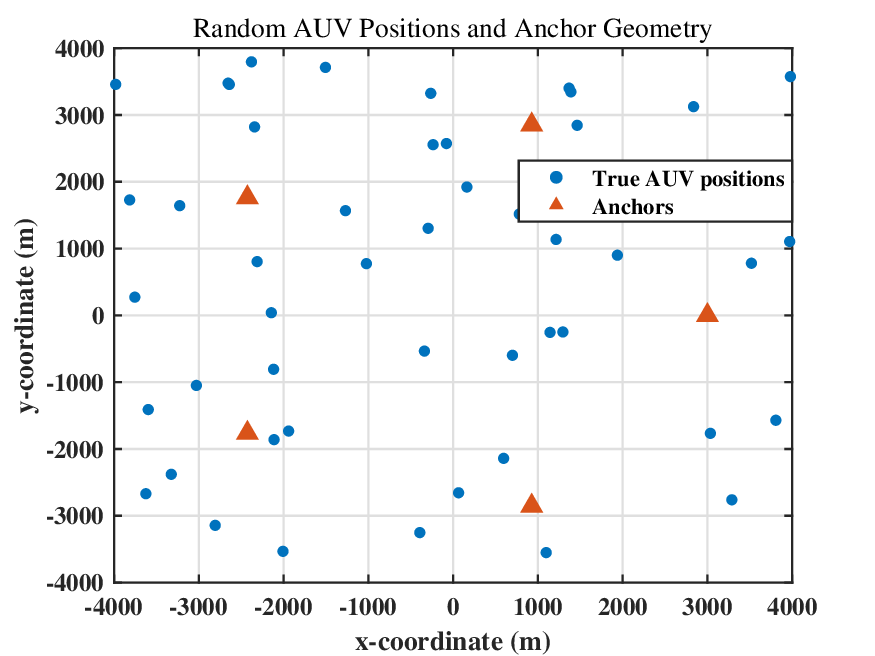}\label{fig05i}}
	\subfloat[]{\includegraphics[width=0.2\textwidth]{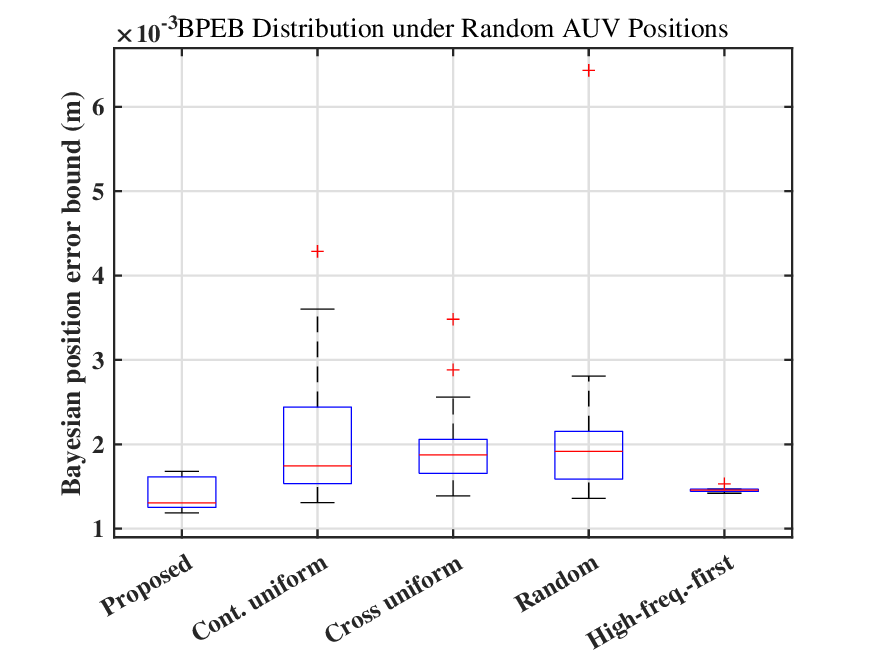}\label{fig05j}}
	\subfloat[]{\includegraphics[width=0.2\textwidth]{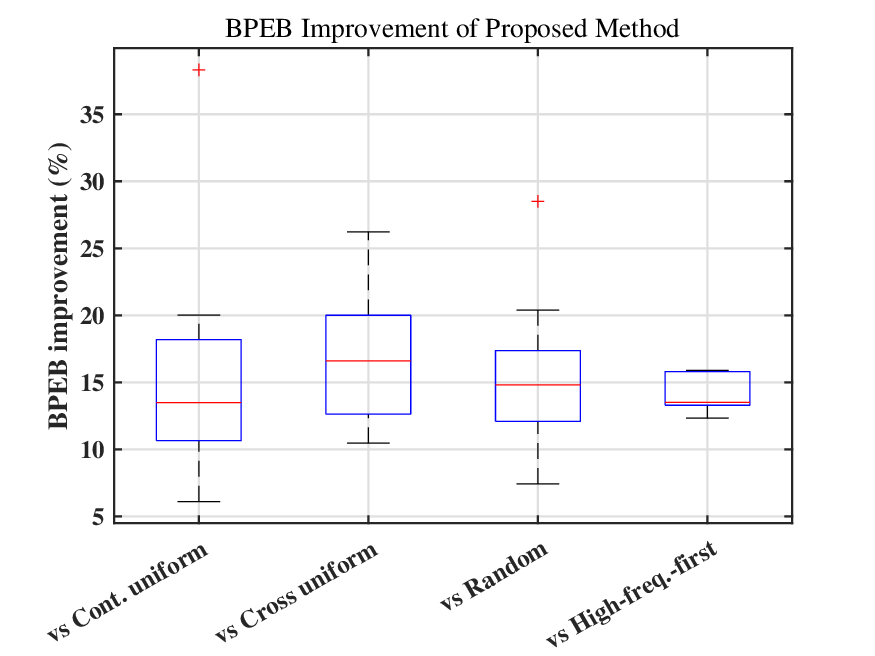}\label{fig05k}}
	\subfloat[]{\includegraphics[width=0.2\textwidth]{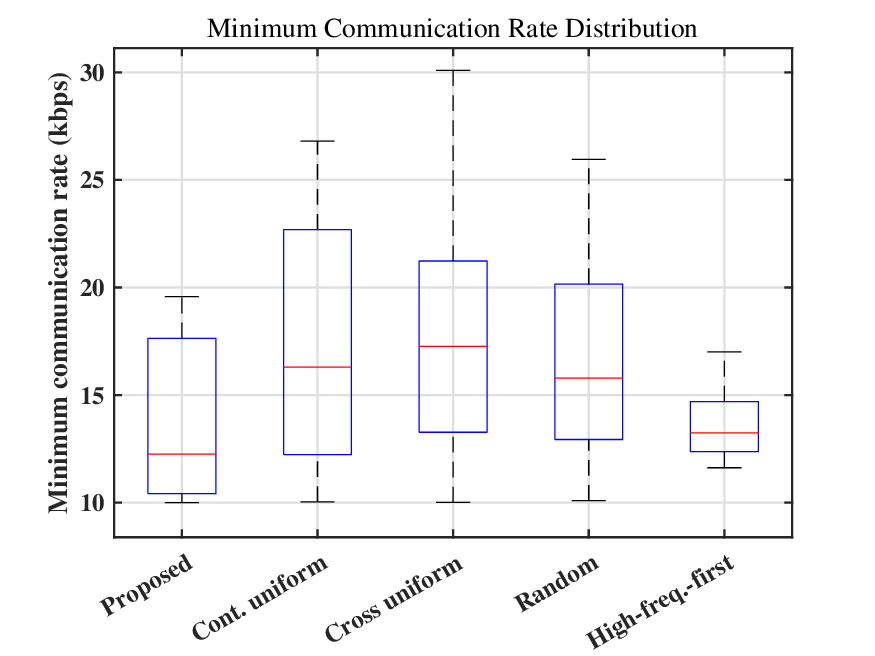}\label{fig05l}}
	\caption{BPEB distributions of different resource-allocation schemes under random AUV positions and Gaussian prior position uncertainty.}\label{fig05}
\end{figure*}

For visualization, the prior-averaged information contribution of anchor $m$ is defined as $I_m^{\mathrm B}\triangleq\operatorname{tr}(\mathbf Q_m)$, where $\mathbf Q_m=\sum_k w_{m,k}\mathbf Q_{m,k}(z_{m,k})$. Fig.~\ref{fig04} shows that both
schemes provide balanced information under the symmetric geometry. Under the asymmetric geometry, uniform allocation produces unequal anchor contributions because of different propagation conditions, whereas the proposed Bayesian allocation compensates for weak links using the prior-averaged matrices $\mathbf Q_{m,k}$. Since $I_m^{\mathrm B}$ only visualizes anchor-level
contributions, the overall localization performance is further evaluated using the complete BPEB.

\subsubsection{BPEB and Communication Feasibility}
To evaluate the robustness of the proposed scheme against prior position uncertainty, Fig.~\ref{fig05} compares the BPEB and communication performance of different subcarrier allocation strategies under symmetric topologies with three, four, and five anchors. For each Monte Carlo realization, the AUV position is randomly generated, while the resource allocation is performed using only a noisy prior position and its Gaussian uncertainty distribution. The expectation in the Bayesian Fisher information matrix is approximated using position samples drawn from this prior distribution. 

As shown in Figs.~\ref{fig05}(b), (f), and (j), the proposed scheme achieves the lowest median BPEB and a relatively concentrated error distribution under all three anchor configurations. The improvements in Figs.~\ref{fig05}(c), (g), and (k) further demonstrate that the proposed scheme reduces the BPEB relative to the four benchmarks for most position realizations. This advantage arises because the proposed allocation jointly exploits the prior-averaged localization information, geometric diversity, frequency-dependent propagation loss, and delay-resolution capability of different subcarriers. In contrast, the two uniform schemes ignore their unequal localization contributions, random allocation cannot systematically exploit frequency diversity, and the high-frequency-first strategy does not jointly balance improved delay resolution against frequency-dependent attenuation and anchor geometry. The improvement does not necessarily increase monotonically with the number of anchors because it also depends on the instantaneous geometry and channel conditions.

\begin{figure*}[!htbp]
	\centering
	\subfloat[]{\includegraphics[width=0.24\textwidth]{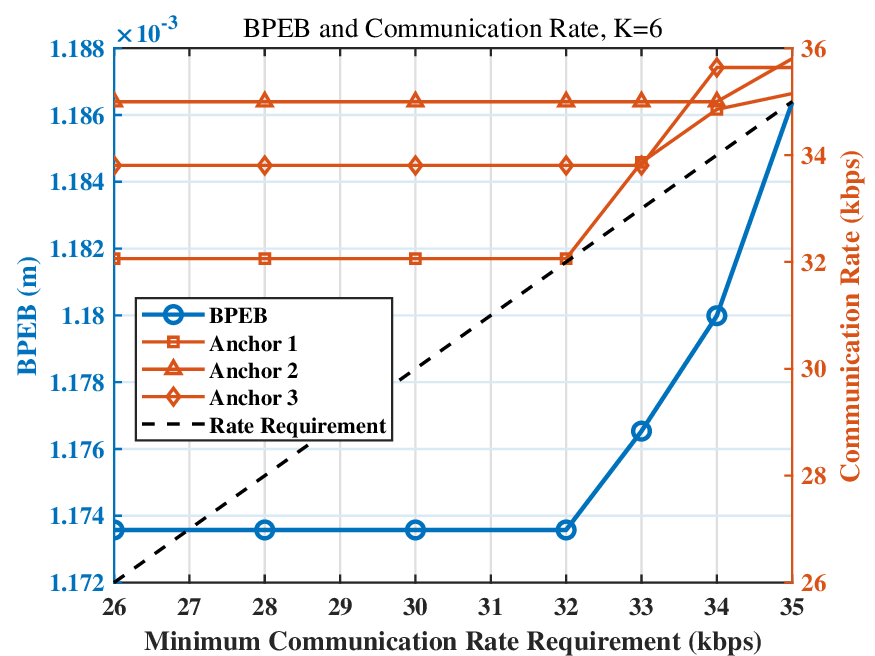}\label{fig06a}}
	\subfloat[]{\includegraphics[width=0.24\textwidth]{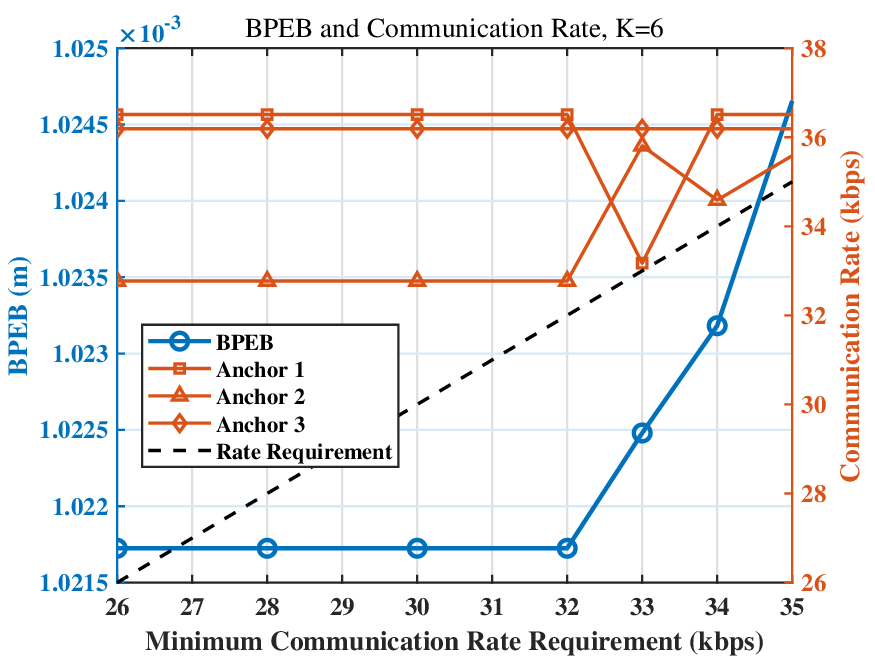}\label{fig06b}}
	\subfloat[]{\includegraphics[width=0.24\textwidth]{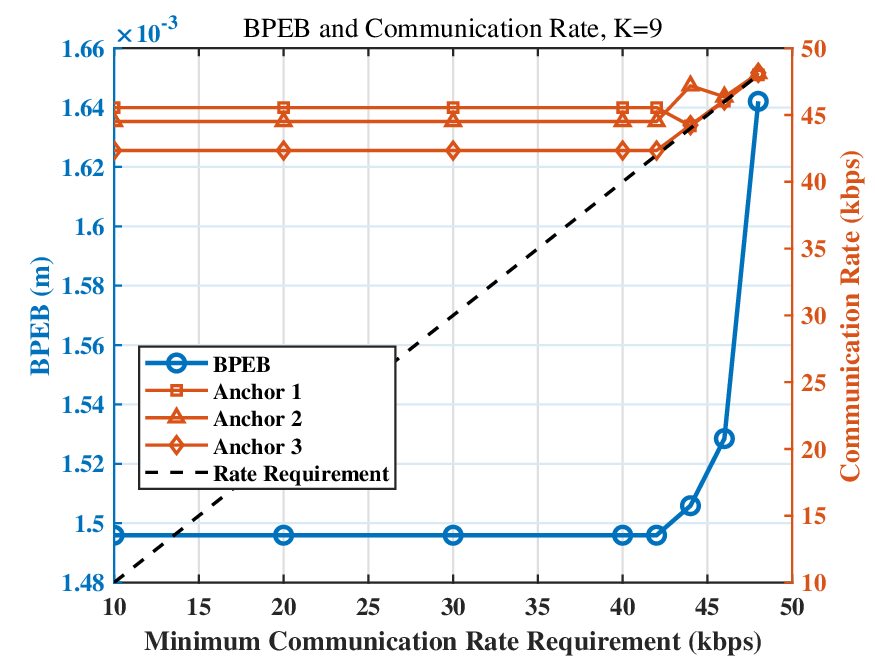}\label{fig06c}}
	\subfloat[]{\includegraphics[width=0.24\textwidth]{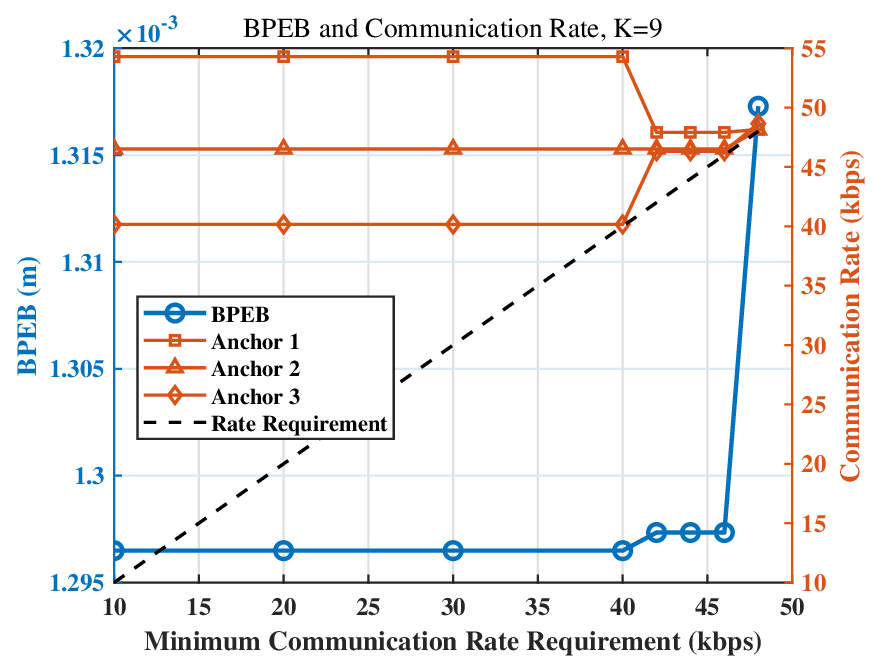}\label{fig06d}}\\
	\subfloat[]{\includegraphics[width=0.24\textwidth]{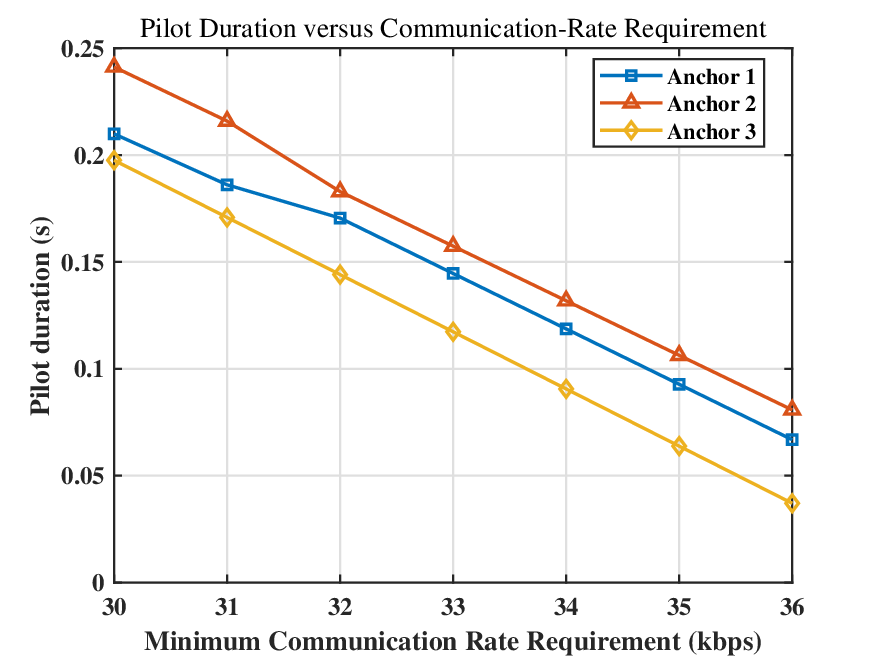}\label{fig06e}}
	\subfloat[]{\includegraphics[width=0.24\textwidth]{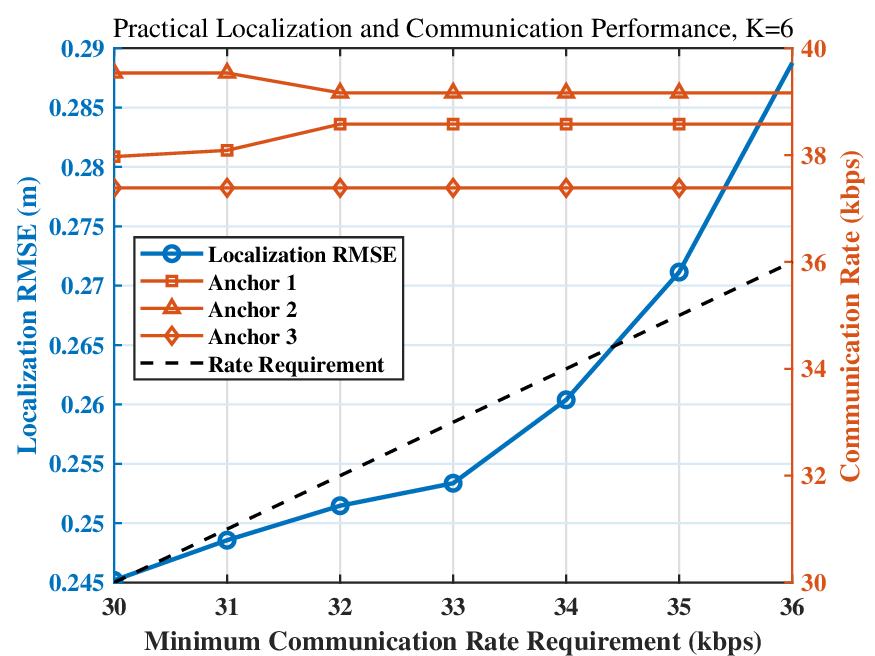}\label{fig06f}}
	\subfloat[]{\includegraphics[width=0.24\textwidth]{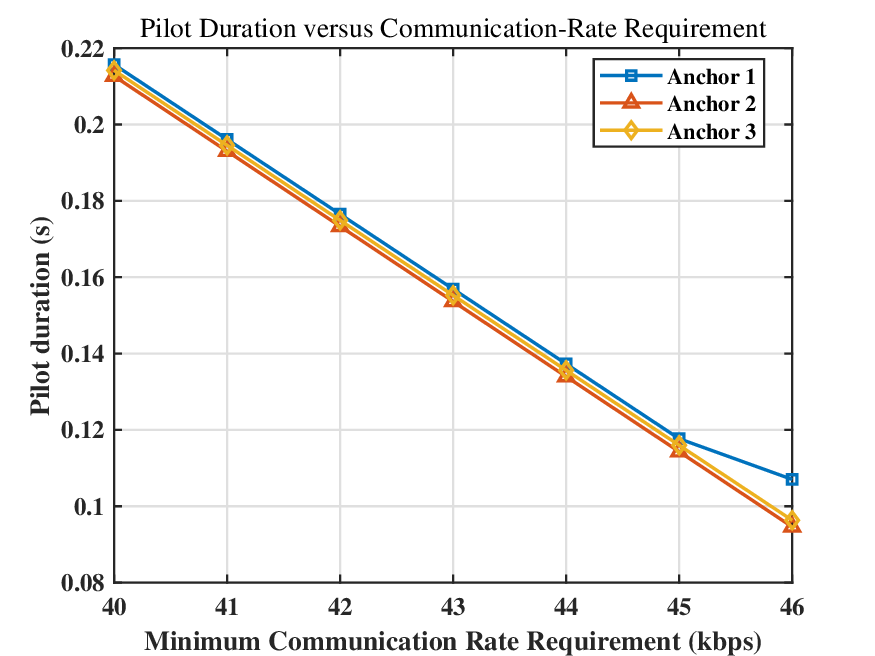}\label{fig06g}}
	\subfloat[]{\includegraphics[width=0.24\textwidth]{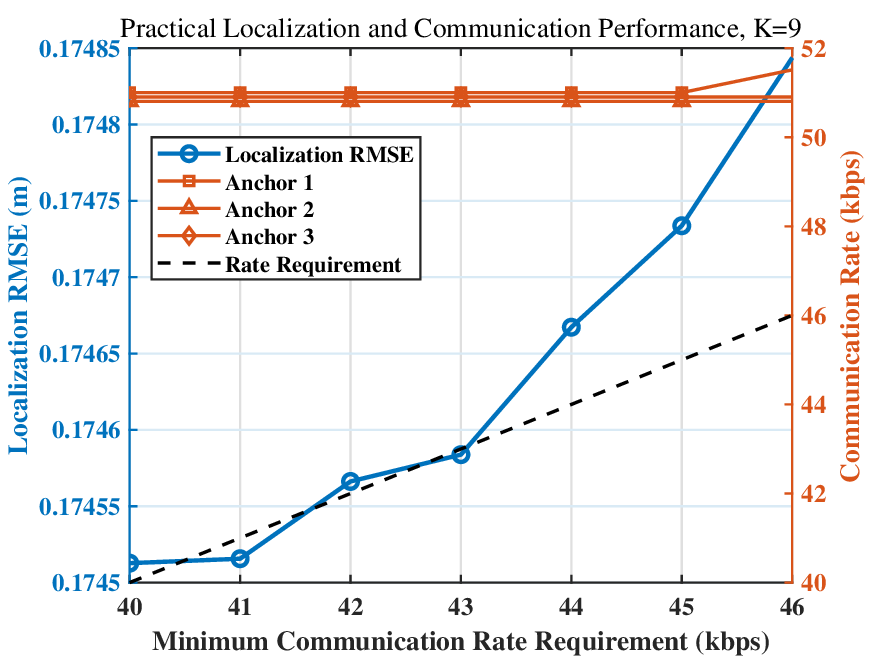}\label{fig06h}}
	\caption{Localization--communication tradeoffs under varying rate
		requirements: (a)--(d) BPEB and data-phase communication rates;
		(e), (g) pilot durations; and (f), (h) horizontal localization
		RMSE and data-phase communication rates.
		Panels (a), (b), (e), and (f) use six subcarriers, whereas
		(c), (d), (g), and (h) use nine subcarriers.
		Panels (a), (c), (e), and (f) correspond to symmetric geometries,
		and (b), (d), (g), and (h) to asymmetric geometries.
		The requirements in (a)--(d) refer to data-phase rates,
		whereas those in (e)--(h) refer to frame-averaged throughput.}\label{fig06}
\end{figure*}

Fig. 6 evaluates the localization–communication tradeoff under different communication-rate requirements. Figs. 6(a) and (b) consider six subcarriers under symmetric and asymmetric anchor geometries, respectively, while Figs. 6(c) and (d) present the corresponding nine-subcarrier cases. In these panels, the BPEB and the individual anchors’ data-phase communication rates are plotted against the minimum data-phase rate requirement.

At relatively low rate requirements, the BPEB remains nearly unchanged, indicating that the communication constraints do not substantially restrict the localization-oriented allocation. As the requirement becomes more stringent, the allocation is adjusted to maintain communication feasibility, accompanied by an increase in the BPEB. This behavior illustrates the loss of localization performance as the feasible resource-allocation set becomes more restrictive. Under the simulated settings, the nine-subcarrier configurations accommodate higher rate requirements but exhibit higher BPEB values than the corresponding six-subcarrier configurations. Thus, these results demonstrate a communication–localization tradeoff rather than a uniform improvement in positioning accuracy with increasing subcarrier count.

Figs.~\ref{fig06}(e)--(h) additionally evaluate link-dependent pilot--datatime allocation with $T_f=1$~s. Here, $R_m$ denotes the data-phase rate, whereas $R_m^{\mathrm{th}}$ specifies the required frame-averaged throughput. For the variable-duration experiments, the pilot power and normalized spectral shape are held fixed, so that the accumulated measurement information scales with the pilot duration relative to the reference duration $T_{\mathrm{pl,ref}}$. The pilot duration and the resulting measurement-information scaling are
\begin{equation}
	\label{eq:pilot_time_scaling}
	T_{\mathrm{pl},m}
	=
	T_f\left(1-\frac{R_m^{\mathrm{th}}}{R_m}\right),
	\qquad
	\mathbf Q_{m,k}^{\mathrm{TDM}}
	=
	\frac{T_{\mathrm{pl},m}}{T_{\mathrm{pl,ref}}}
	\mathbf Q_{m,k}^{\mathrm{ref}},
\end{equation}
where $T_{\mathrm{pl},m}\geq0.02$~s and $T_{\mathrm{pl,ref}}=0.2$~s.

As shown in Figs.~\ref{fig06}(e) and~\ref{fig06}(g), increasing the required frame-averaged throughput reduces the time available for localization pilots within the fixed frame. The pilot durations differ among anchors because their data-phase rates depend on the allocated subcarriers, power, and propagation conditions.

Figs.~\ref{fig06}(f) and~\ref{fig06}(h) show the corresponding LS localization RMSE and data-phase communication rates. The shorter pilot durations reduce the accumulated ranging information and increase the simulated TOA uncertainty, resulting in the observed increase in localization RMSE. The plotted communication rates represent
data-phase rates, whereas the effective frame-averaged rates equal the prescribed requirements under the adopted time-split rule. These results illustrate the impact of communication time requirements on localization performance.

\subsubsection{Practical Evaluation}
\begin{figure*}[!htbp]
	\centering
	\subfloat[]{\includegraphics[width=0.15\textwidth]{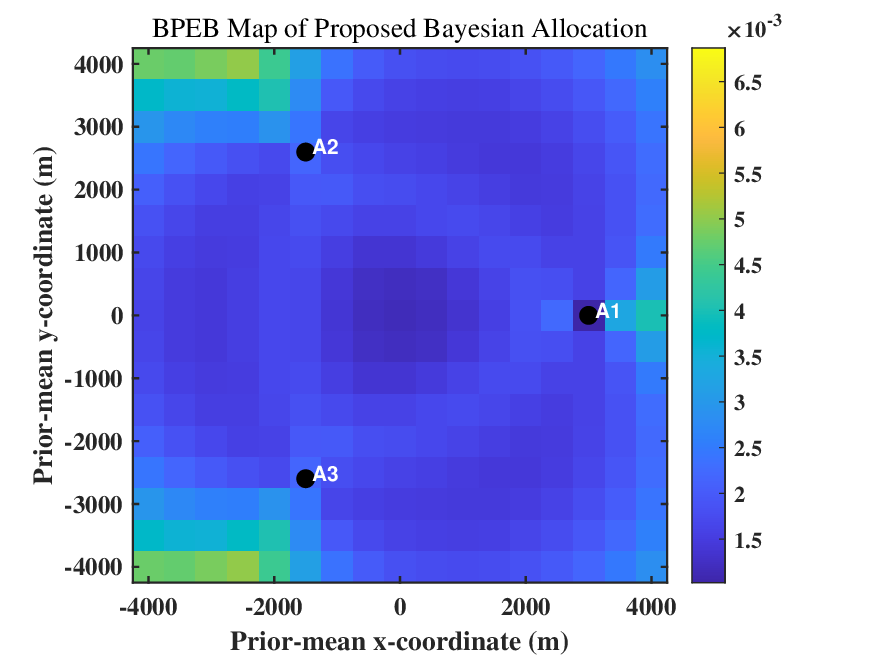}\label{fig07a}}
	\subfloat[]{\includegraphics[width=0.15\textwidth]{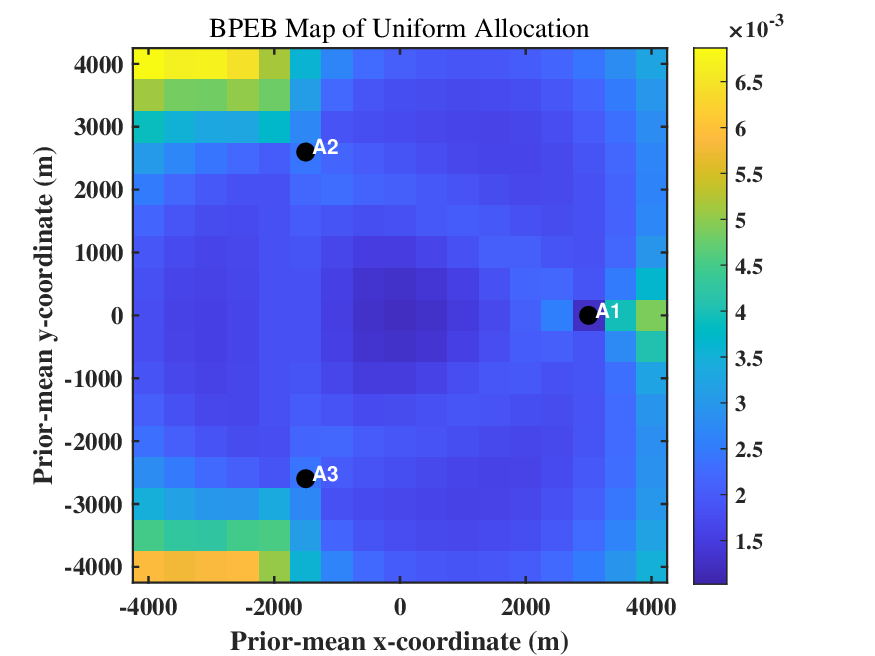}\label{fig07b}}
	\subfloat[]{\includegraphics[width=0.15\textwidth]{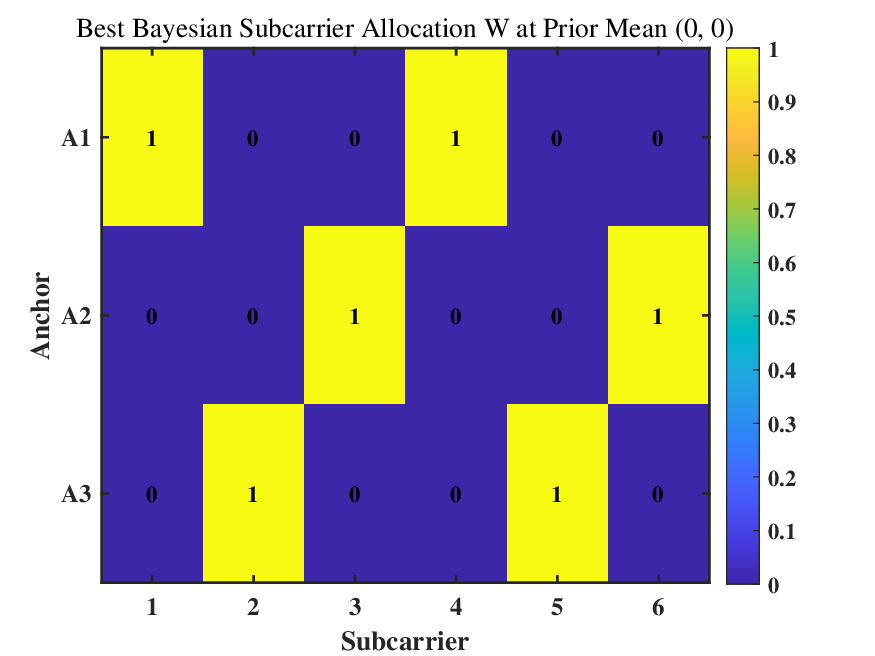}\label{fig07c}}
	\subfloat[]{\includegraphics[width=0.15\textwidth]{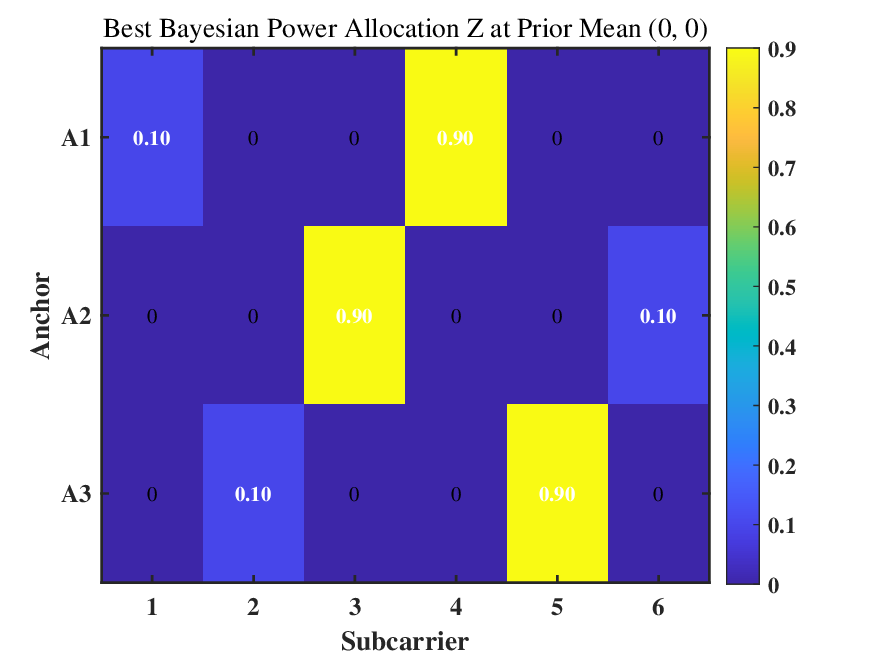}\label{fig07d}}
	\subfloat[]{\includegraphics[width=0.15\textwidth]{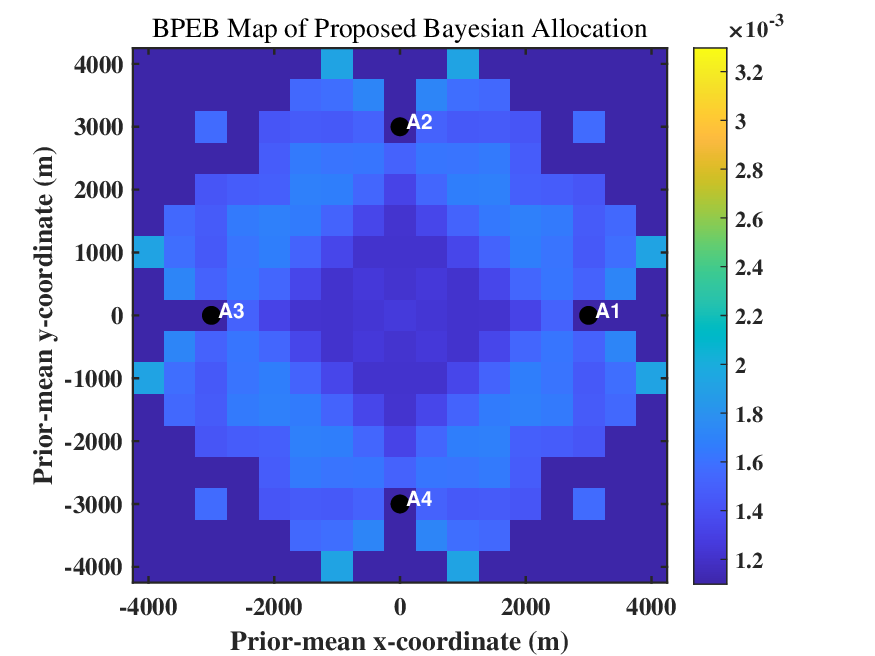}\label{fig07e}}
	\subfloat[]{\includegraphics[width=0.15\textwidth]{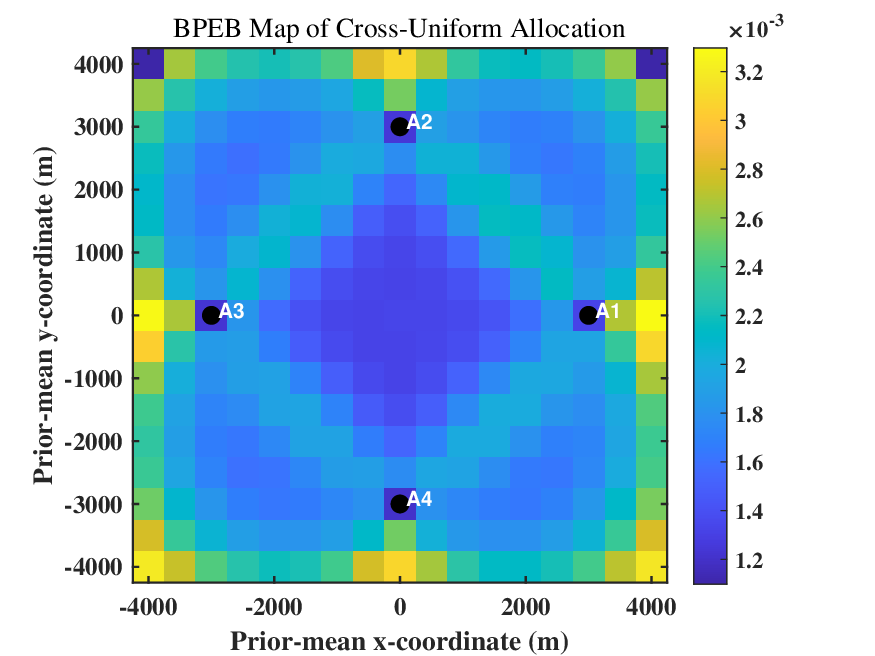}\label{fig07f}}\\
	\subfloat[]{\includegraphics[width=0.15\textwidth]{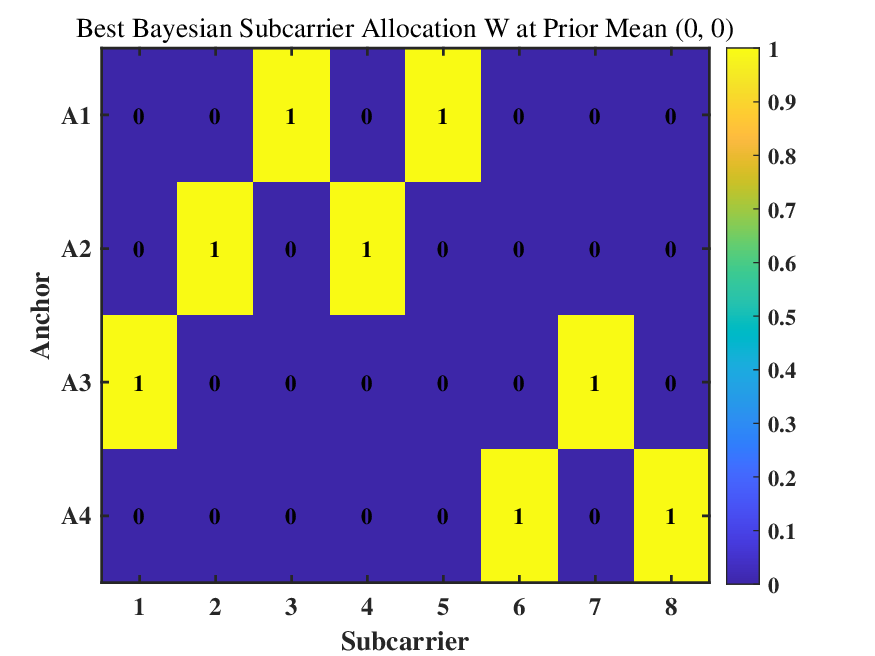}\label{fig07g}}
	\subfloat[]{\includegraphics[width=0.15\textwidth]{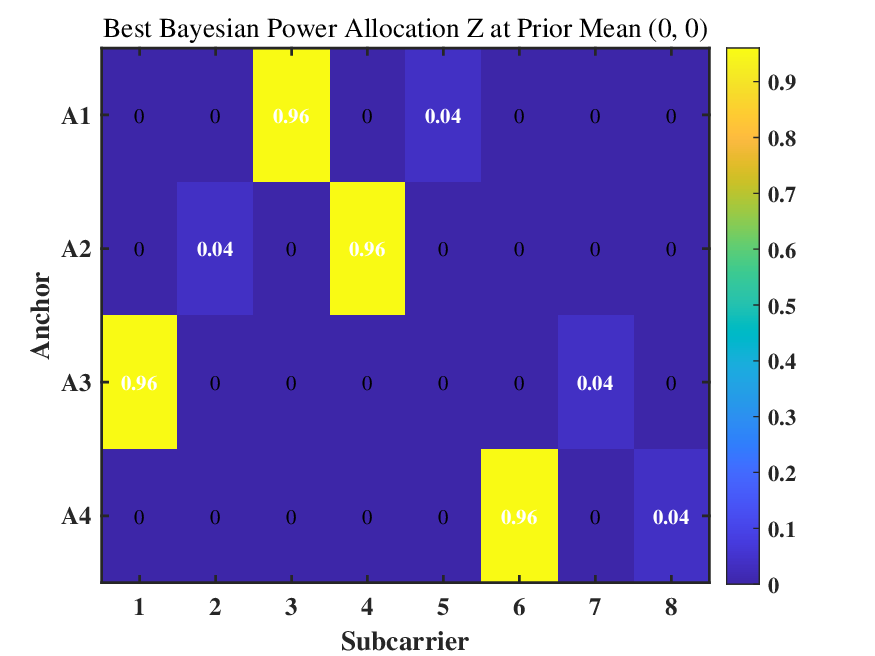}\label{fig07h}}
	\subfloat[]{\includegraphics[width=0.15\textwidth]{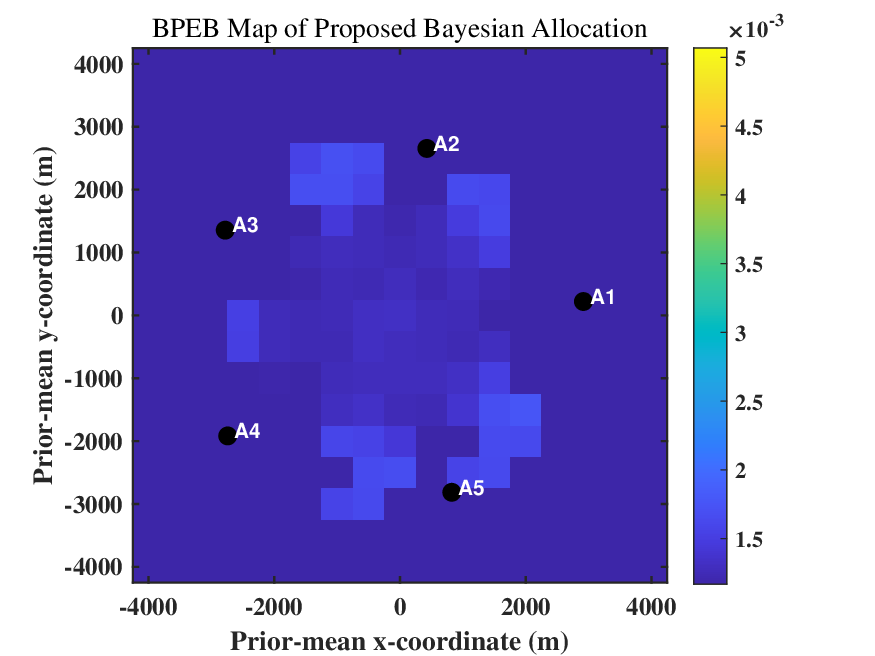}\label{fig07i}}
	\subfloat[]{\includegraphics[width=0.15\textwidth]{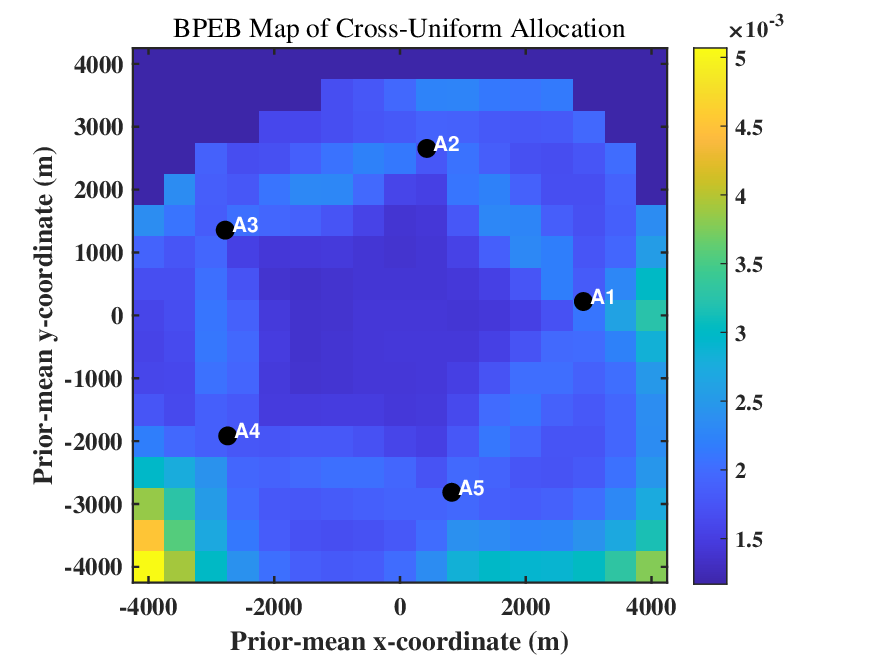}\label{fig07j}}
	\subfloat[]{\includegraphics[width=0.15\textwidth]{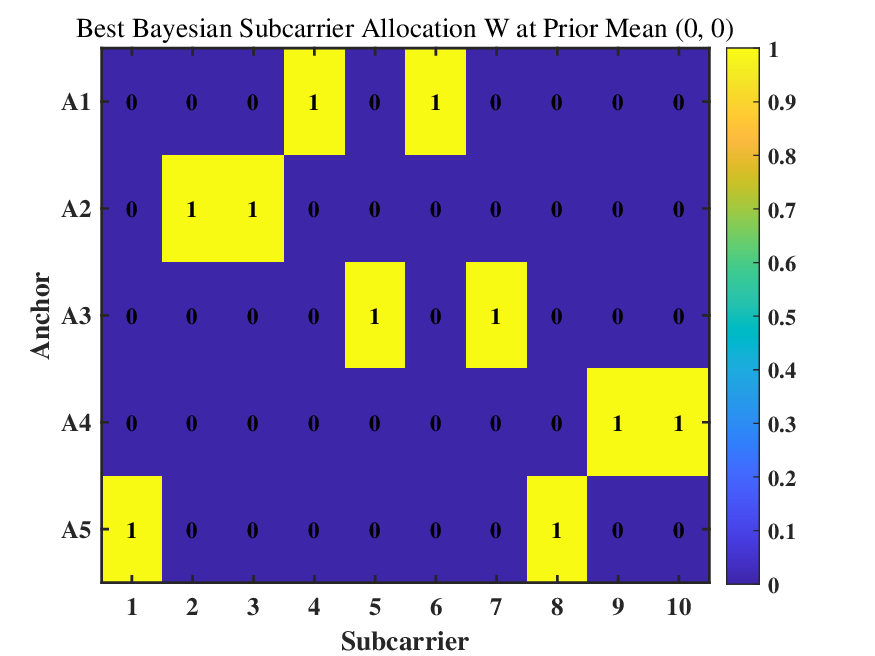}\label{fig07k}}
	\subfloat[]{\includegraphics[width=0.15\textwidth]{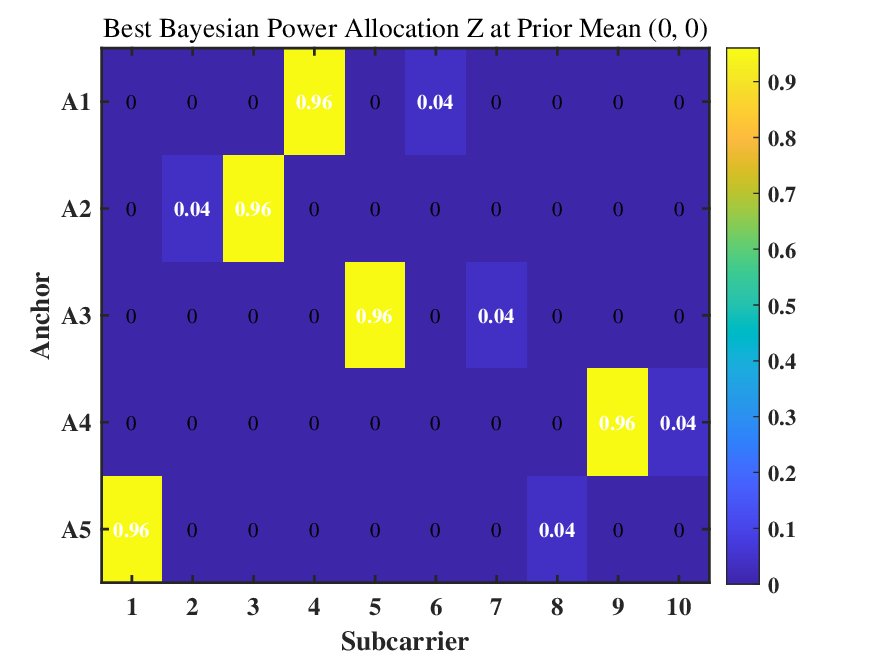}\label{fig07l}}
	\caption{BPEB maps over prior mean horizontal positions and
		resource allocations at $\hat{\mathbf p}_0=(0,0)^T$:
		(a)--(d) three symmetric anchors;
		(e)--(h) four symmetric anchors;
		(i)--(l) five asymmetric anchors.
		The AUV depth is known and fixed.}\label{fig07}
\end{figure*}
\begin{figure*}[!htbp]
	\centering
	\subfloat[]{\includegraphics[width=0.15\textwidth]{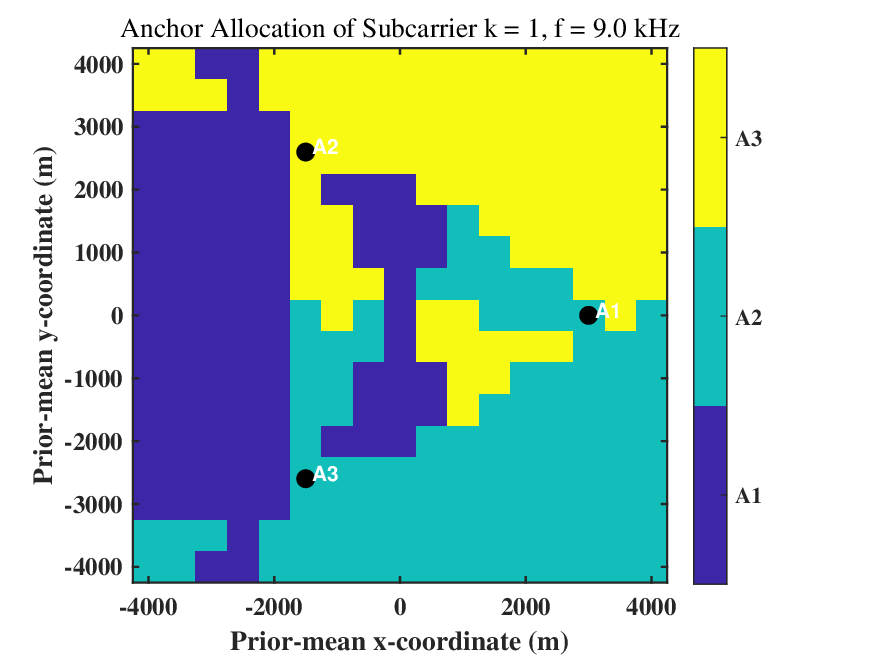}\label{fig08a}}
	\subfloat[]{\includegraphics[width=0.15\textwidth]{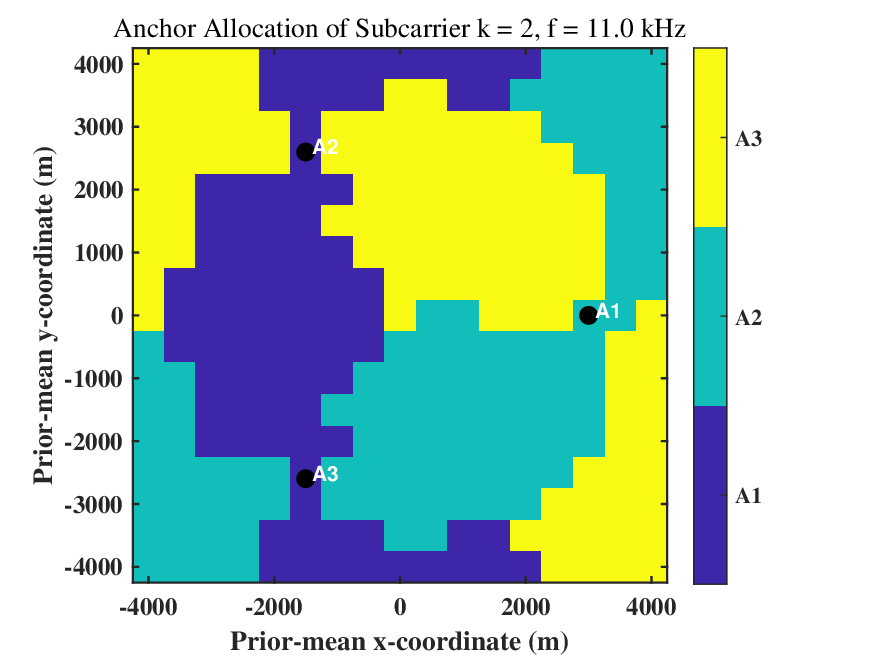}\label{fig08b}}
	\subfloat[]{\includegraphics[width=0.15\textwidth]{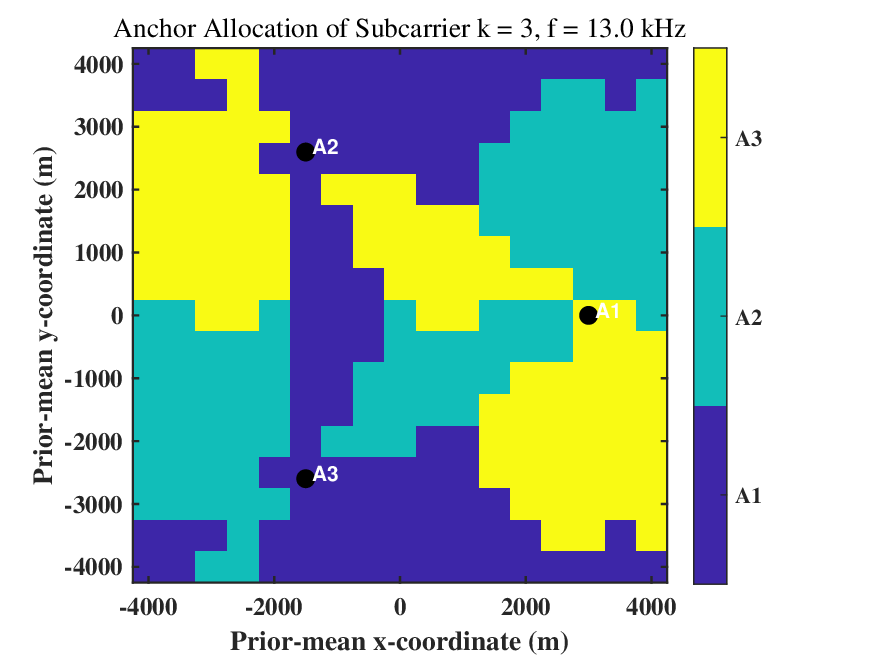}\label{fig08c}}
	\subfloat[]{\includegraphics[width=0.15\textwidth]{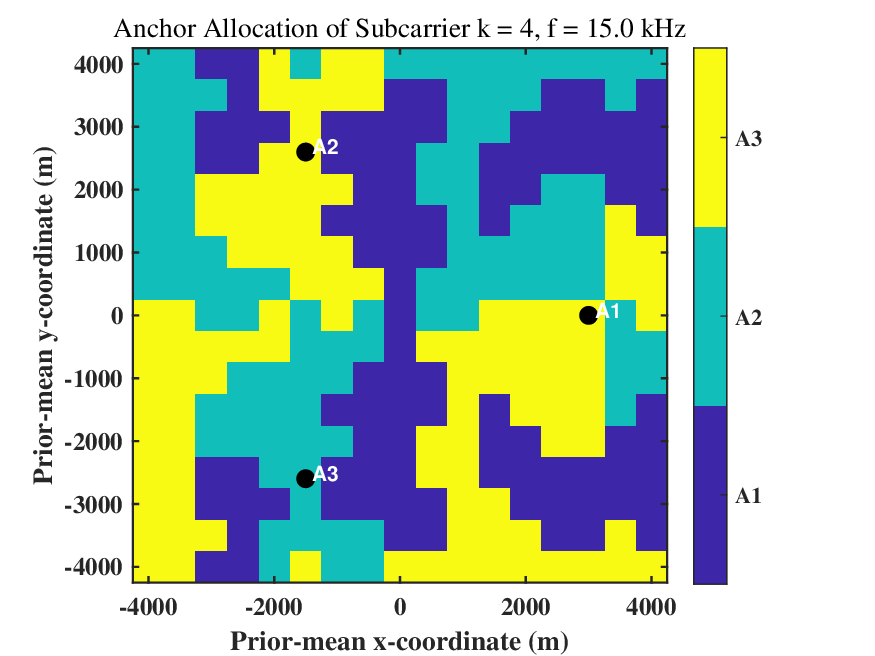}\label{fig08d}}
	\subfloat[]{\includegraphics[width=0.15\textwidth]{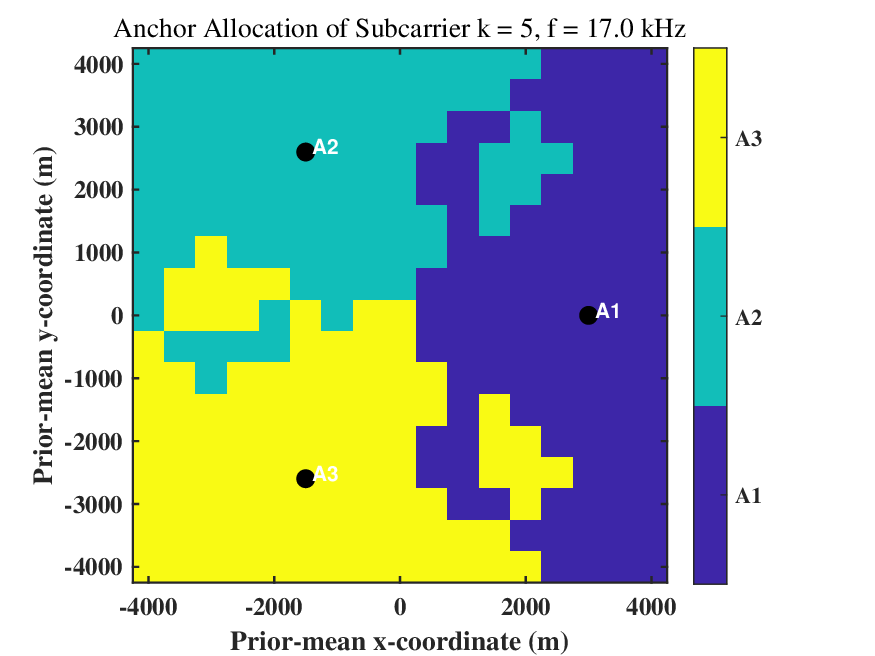}\label{fig08e}}
	\subfloat[]{\includegraphics[width=0.15\textwidth]{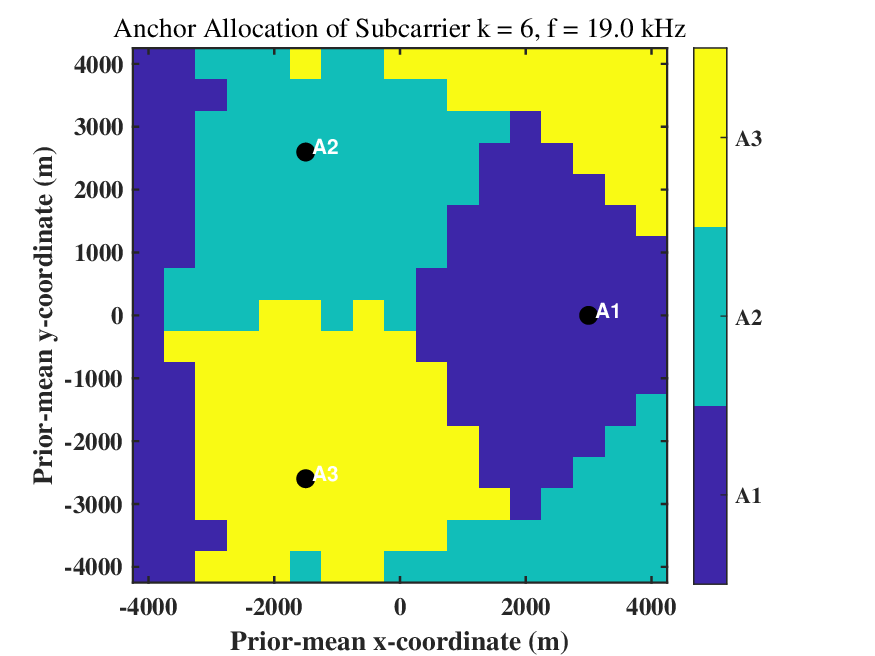}\label{fig08f}}
	\caption{Subcarrier-to-anchor assignments over prior mean
		horizontal positions for the symmetric three-anchor topology:
		(a) $f_1=9$ kHz;
		(b) $f_2=11$ kHz;
		(c) $f_3=13$ kHz;
		(d) $f_4=15$ kHz;
		(e) $f_5=17$ kHz; and
		(f) $f_6=19$ kHz.
		Colors identify the anchor assigned to each subcarrier.}\label{fig08}
\end{figure*}
Fig.~\ref{fig07} presents the BPEB maps over different prior mean horizontal positions, together with the resource allocations obtained at $\hat{\mathbf p}_0=(0,0)^T$. At each grid point, resource allocation and BPEB evaluation use the Gaussian position prior centered at that point, with the AUV depth held fixed. The spatial variation in BPEB reflects changes in the prior-averaged anchor--AUV geometry and propagation conditions. Nevertheless, the proposed allocation strategy consistently maintains favorable localization accuracy across the entire service area. As the number of anchors increases, the overall BPEB decreases because additional anchors provide more localization information and geometric diversity. The optimal allocation results further show that the proposed method adaptively distributes subcarriers and power resources according to the localization contribution of each anchor, thereby improving the utilization of available frequency resources. The performance gain is particularly noticeable in the asymmetric topology, where resource allocation plays a more important role in compensating for geometric imbalance.

Fig.~\ref{fig08} presents the optimal subcarrier allocation results for the symmetric three-anchor topology. It can be observed that lower-frequency subcarriers are preferentially assigned to distant anchors, whereas higher-frequency subcarriers tend to be allocated to nearby anchors, which is consistent with Theorem 1. This allocation pattern arises from the frequency-dependent propagation loss and acoustic absorption characteristics of underwater channels. Since localization information varies across both distance and frequency, the proposed method adaptively assigns subcarriers according to their localization contributions, thereby improving the overall positioning performance. The symmetric patterns observed are mainly caused by the symmetric anchor deployment. Since the proposed allocation adapts the frequency resources according to the position prior centered at each grid point, the BPEB distribution exhibits local distortions compared with uniform allocation, indicating that the localization information distribution is jointly affected by both anchor geometry and frequency allocation.

\begin{figure}[!htbp]
	\centering
	\subfloat[]{\includegraphics[width=0.24\textwidth]{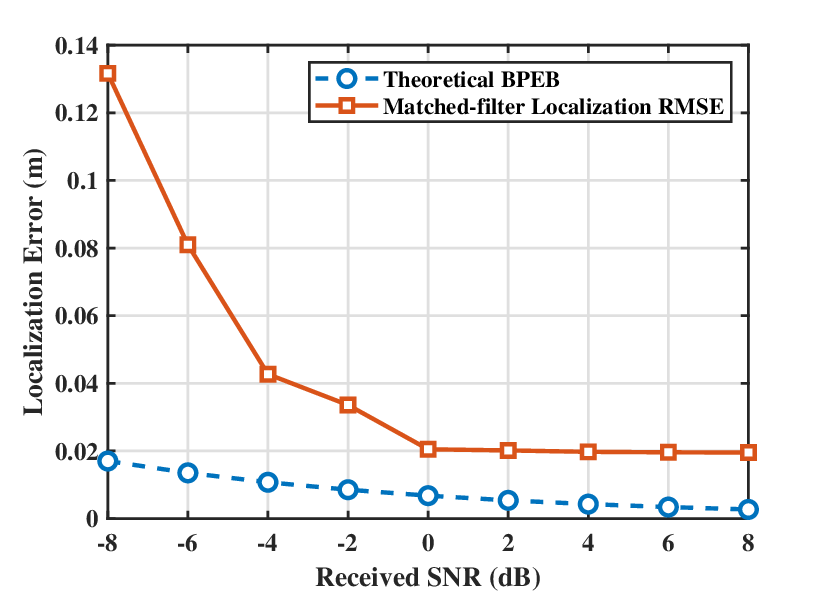}\label{fig09a}}
	\subfloat[]{\includegraphics[width=0.24\textwidth]{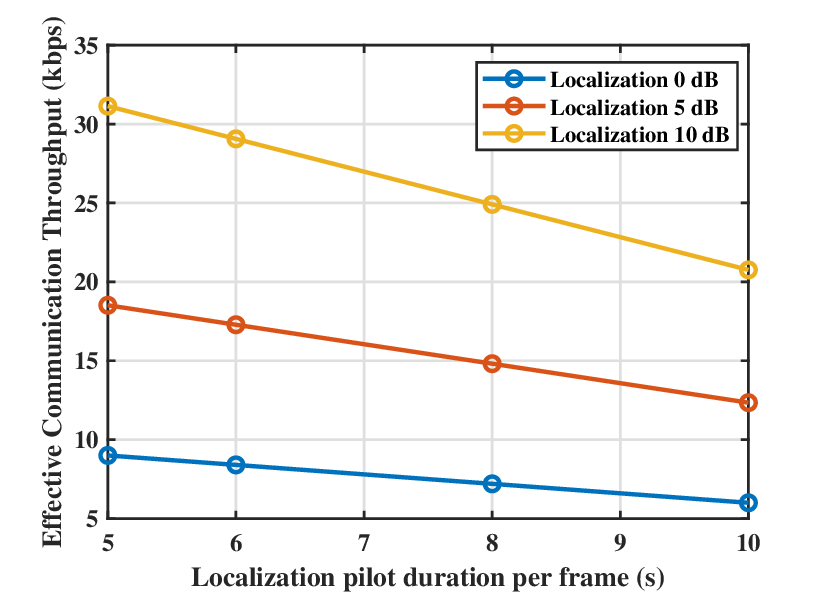}\label{fig09b}}
	\caption{Localization and communication performance:
		(a) theoretical BPEB and matched-filter-based localization
		RMSE versus received SNR;
		(b) effective communication throughput versus localization
		pilot duration at received SNRs of 0, 5, and 10 dB,
		with a frame duration of 20 s.}\label{fig09}
\end{figure}

Fig.~\ref{fig09a} compares the BPEB-based localization error with the matched-filter-based localization result under different received SNRs. As the received SNR increases, both localization errors decrease because more accurate TOA estimates can be obtained from the received signals. The matched-filter-based localization error closely follows the theoretical BPEB trend, confirming the validity of the derived performance bound. A small but non-negligible gap remains between the two curves across the entire SNR range. This gap is mainly caused by systematic errors in the localization model, including geometric approximation errors, discretization effects in TOA estimation, and the nonlinearity of the position estimation process. Consequently, the practical localization performance approaches but does not exactly attain the theoretical BPEB.

\subsection{Communication Performance Evaluation}
To evaluate the communication performance of the adopted multi-carrier ISAC framework, Fig. 9(b) presents the effective communication throughput versus the localization pilot duration at different received SNRs. Consistent with the signal model in Section II, localization pilots and communication data occupy separate time intervals within each transmission frame while sharing the allocated subcarriers. Let \(T_{\mathrm{frame}}\) denote the frame duration and \(T_{\mathrm{loc}}\) denote the duration reserved for localization pilots. The remaining duration, \(T_{\mathrm{frame}}-T_{\mathrm{loc}}\), is available for data transmission. Accordingly, the effective communication throughput is obtained by multiplying the data-phase communication rate by the data transmission fraction \(1-T_{\mathrm{loc}}/T_{\mathrm{frame}}\).

As shown in Fig.~\ref{fig09}(b), the effective communication throughput decreases approximately linearly as the localization pilot duration increases at each SNR. With the data-phase resource allocation and SNR held fixed, this reduction results from the shorter time available for communication rather than a deterioration in channel quality. Moreover, a higher received SNR supports a higher data-phase communication rate and therefore yields greater effective throughput for the same pilot duration. These results quantify the communication cost of reserving time for localization pilots within the adopted TDM frame structure and highlight the need to select the pilot duration according to both localization and communication requirements.

\subsection{Computational Complexity Analysis}
\begin{table}[!htbp]
	\centering
	\caption{Computational Complexity Evaluation for different frequency allocation methods (time ms).}
	\label{tab02}
	\begin{tabular}{lccc}
		\hline
		Method \& Subcarriers & 6 & 9 & 12 \\
		\hline
		Pre-processing       & 4926.6 & 4930.1 & 4943.4 \\
		\hline
		Proposed             & 3.07 & 7.69 & 9.89 \\
		Random               & 2.19 & 2.33 & 2.61 \\
		High-frequency-first & 0.03 & 0.05 & 0.06 \\
		\hline
	\end{tabular}
\end{table}
Finally, we compare the computational complexity among different frequency allocation methods as shown in Table~\ref{tab02}. As expected, the proposed method requires a higher computational cost than the heuristic baselines because it jointly considers localization information and communication constraints during the allocation process. Nevertheless, the average runtime remains below 10 ms even when the number of subcarriers increases to 12, demonstrating the practical feasibility of the proposed algorithm.

In contrast, the high-frequency-first scheme exhibits the lowest complexity since it only performs a simple sorting operation without iterative optimization. The random allocation method requires slightly more computation due to repeated random assignments and performance evaluation. Although the proposed method incurs additional computational overhead, the runtime growth remains moderate as the problem size increases. Combined with the substantial localization performance gains observed in previous experiments, the results indicate that the proposed algorithm achieves an effective balance between computational complexity and positioning accuracy.

\section{Conclusion}
This work investigated localization-oriented subcarrier allocation for multi-carrier underwater ISAC systems. A unified communication-localization framework was developed, in which a closed-form TOA CRB was derived under frequency-selective channels and colored interference. To avoid assuming perfect knowledge of the AUV position during resource allocation, a Bayesian positioning framework was established by incorporating a Gaussian position prior and averaging the measurement Fisher information over the position uncertainty. Accordingly, a prior-averaged subcarrier-level localization information metric was introduced to enable distance-frequency matching and balanced information allocation.

The resource allocation problem was formulated as an A-optimal optimization with communication constraints, and a low-complexity greedy algorithm with power refinement was proposed. Simulation results demonstrate improved
localization accuracy over conventional schemes while maintaining communication performance. Future work will consider mobile scenarios and robust allocation under channel uncertainties.

\bibliographystyle{IEEEtran}
\bibliography{IEEEabrv,draft_hw}

\end{document}